\documentclass[prb,preprint,amsmath,amssymb,floatfix,footinbib,bibnotes,longbibliography]{revtex4-1}

\pdfoutput=1

\usepackage{soul}

\usepackage[colorlinks=true,citecolor=blue,linkcolor=magenta]{hyperref}
\usepackage{graphicx}
\usepackage{changepage}
\usepackage{fancyhdr}
\usepackage{amsthm, amssymb}
\usepackage{float}
\usepackage{array}
\usepackage[usenames,dvipsnames]{color}

\usepackage{rotating}
\usepackage{hyperref}
\usepackage{bm}
\usepackage{color}

\def \be{\begin{equation}}
\def \ee{\end{equation}}
\def \bea{\begin{eqnarray}}
\def \eea{\end{eqnarray}}

\def \ba{\begin{align}}
\def \ea{\end{align}}

\newcommand {\apgt} {\ {\raise-.5ex\hbox{$\buildrel>\over\sim$}}\ }
\newcommand {\aplt} {\ {\raise-.5ex\hbox{$\buildrel<\over\sim$}}\ }

\begin{document}

\title{High Temperature Superconductivity in the Cuprates: Materials, Phenomena and a Mechanism}

\author{Sumilan Banerjee$^{1}$, Chandan Dasgupta$^{1,2}$, Subroto Mukerjee$^{1}$, TV Ramakrishnan$^{1,3}$ and Kingshuk Sarkar$^{1,4}$ \\
\small \em $^1$Department of Physics, Indian Institute of Science, Bangalore 560012, India\\
$^{2}$ International Centre for Theoretical Sciences, Tata Institute of Fundamental Research, Bengaluru 560089, India\\
$^{3}$ Department of Physics, Banaras Hindu University, Varanasi 221005, India\\
$^{4}$ Physics Department, Ben Gurion University, Beer Sheva 84105, Israel }

\begin{abstract}
Superconductivity in the cuprates, discovered in the late 1980’s and occurring at unprecedentedly high temperatures (up to about 140K) in about thirty chemically distinct families, continues to be a major problem in physics. In this article, after a brief introduction of these square planar materials with weak interlayer coupling, we mention some of the salient electronic properties of hole doped cuprates such as the pseudogap phase and the Fermi arc . We then outline a phenomenological, Ginzburg Landau like theory developed by some of us for the emergent d-wave symmetry superconductivity in these materials, and confronted successfully with a large amount of experimental information. A more recent application of the approach to fluctuation diamagnetism and to the anomalously large Nernst effect is also discussed.
\end{abstract}
\maketitle

\section{Introduction}
Superconductivity was discovered more than a century ago, and seemed to be a very low temperature phenomenon confined to temperatures below about 20K. Therefore the discovery of high temperature superconductivity in a family of cuprates around 1987 or so took  the Physics world by surprise, not only because the temperatures ($T_c$) for transition to superconductivity are high (by values prevalent till then in the field; the cuprate $T_c$'s range from about 30K to about 140K), but also because these materials have unusual physical properties. Major unexpected experimental discoveries continue to be made, and there is no single broadly accepted theoretical framework in which the wide range of observed properties can be understood and calculated.  In this article, we will outline some of the broad materials characteristics of these systems, some of their basic properties, and outline a novel phenomenological approach which is used to compute a wide range of experimental properties. 

The article is organized as follows. Section \ref{sec.materials} of this article is an introduction to the family of cuprate materials with a description of their structure and a \emph{minimal} electronic model appropriate to the cuprates. A long section (Section \ref{sec.cupratephenomena}) describes the phase diagram of the hole doped cuprates, which in addition to antiferromagnetically ordered insulating phase and a superconducting region, hosts `pseudogap' and strange metal regions. The unusual electronic properties of these phases are also described. (This is what makes the section long). Section \ref{sec.GLTheory} introduces our phenomenological, superficially Ginzburg-Landau-like approach to the system. In Section \ref{sec.extensions}, we outline further extensions of this phenomenological theory to describe transport properties (especially the unusual Nernst or thermo-magneto-electric effect) and electron spectra as measured in ARPES (Angle Resolved Photo Emission Spectroscopy). Section \ref{Sec.Results} mentions some of the major results obtained from the theory for a broad range of equilibrium and nonequilibrium phenomena, above and below $T_c$ and for a range of doping $x$. The concluding section (Section \ref{sec.conclusion}) mentions some future directions.

\section{Materials, structure and minimal electronic model} \label{sec.materials}
\begin{figure} [hbt]
\begin{center}
\includegraphics[width=6cm]{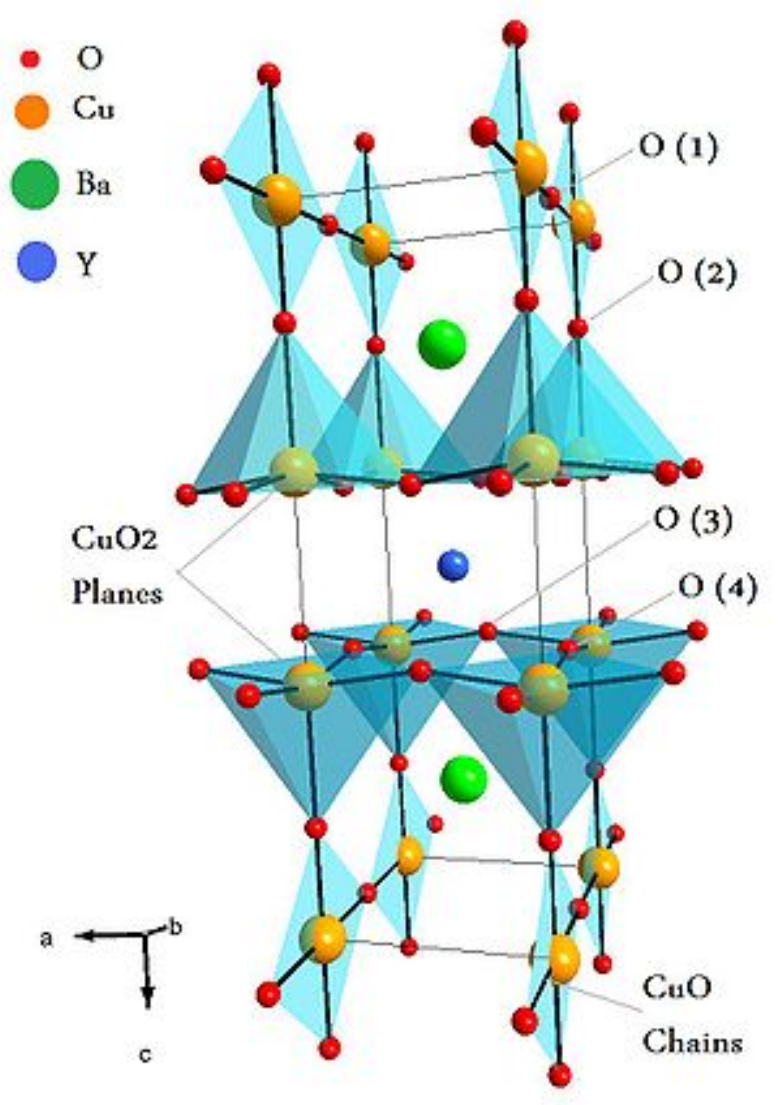}
\end{center}
\caption{\footnotesize{Orthorhombic unit cell of $\mathrm{YBa_2Cu_3O_{7-x}}$
(YBCO or Y123) with $a=3.82~\AA$ along $[100]$ and $b=3.89~\AA$ along $[010]$. The $c$ axis dimension is around 12 $\AA$.}}
\label{fig.YBCO_Crystal}
\end{figure}

 The cuprates , e.g. $\mathrm{La_{2-x} Sr_xCuO_4}$ , $\mathrm{YBa_2Cu_3O_{7-x}}$ , $\mathrm{Bi_2Sr_2Ca_{n-1}Cu_nO_{2n+4+x}}$ ( n=1,2,3)  can be thought of as being made up of highly distorted corner sharing octahedra derivable from the $\mathrm{ABO_3}$ perovskite family.  The $\mathrm{O^{2-}}$ ions are at the corners of the octahedron, the $\mathrm{Cu^{2+}}$  ions at the centre, and the transition metal or rare-earth ions in the space between the octahedra. The Jahn-Teller distortion due the $\mathrm{Cu^{2+}}$ ions is very large; the plane of the octahedron is split in two, with a large splitting. (Fig.\ref{fig.YBCO_Crystal} shows the structure of $\mathrm{YBa_2Cu_3O_{7-x}}$).  A good approximation for the electronic properties of these systems is to regard the materials as consisting of Cu-O planes; the Cu and O ions are arranged in squares in the plane, with O ions at the midpoint of the nearest Cu-Cu line.  The low energy electronic degrees of freedom arise from the unfilled $d$ shell electrons of the $\mathrm{Cu^{2+}}$ ions which are in the $d^9$ configuration. The Cu ions are at the vertices of a square whose length in most cuprates is about $3.9\AA$.  Crudely, at the single site level, the occupation of the nine $d$ electrons of the $\mathrm{Cu^{2+}}$ ion can be described as follows. The tenfold atomically degenerate $d$ level is split into $t_{2g}$ and $e_g$ configurations, with the former lying lower in energy. Its six states are fully occupied by $d$ electrons.  That they do not take part in any low energy electron processes, and are strongly bonded with $\mathrm{O^{2-}}$  electrons, is clear from the fact that the relevant bond lengths, reflected in the Cu-Cu distances, are about the same in almost all cuprates. The twofold degenerate $e_g$ level is strongly split in two by the Jahn-Teller distortion mentioned above; the lower one is fully occupied and the higher one is half occupied in the $d^9$ configuration. 
 
 For large local Coulomb repulsion or Mott Hubbard $U$ (as is widely believed to be the case in cuprates from both theoretical and experimental evidence ) the fluctuations in the number of local $d$ electrons is strongly suppressed, and the material is an insulator for exactly one $e_g$ electron per site (Mott insulator). If the cuprate is hole doped (for example, the trivalent La is  partially substituted by the divalent Sr) both naively, and from a more sophisticated model which includes the oxygen $p$ electrons, it is clear that the system can be thought of electronically  as consisting (on the average) of $(1-x)$  electrons per unit cell where a fraction $x$ of the trivalent La ions is substituted by Sr ions. These electrons move on a square lattice, and in a tight binding picture can be regarded as hopping from one site to another on a square planar lattice with hopping amplitudes $t,~t'$ etc.. There is an energy penalty of $U$ for double occupation of a particular site by electrons. There is an amplitude $\lambda t$ ($\lambda<<1$) for electron hopping perpendicular to the plane. We will neglect this in what follows. Based on the above picture, a one-band Hubbard model on a square lattice is believed to be a reasonable microscopic description for most of the universal aspects of the cuprate phase diagram. The most widely used values for these parameters which define the electron dynamics of the model are, $t=300$ meV, $t'/t=-1/4$ \cite{AParamekanti1-6} and $U=2-4$ eV \cite{AFujimori1-2}. The large value of $U$ ($U\gg 4t$, the half bandwidth) implies that the material is a strongly correlated metal in which the electron of the Cu hops from site to site, avoiding other $d$ electrons. In the large $U$ limit the one-band model Hubbard model reduces to the well-known $t-J$ model, 
\begin{eqnarray}
\hat{\mathcal{H}}_{t-J}&=&-\sum_{<ij>,\sigma}\mathbb{P}_G\left(t_{ij}a^\dagger_{i\sigma}a_{j\sigma}+\mathrm{h.c.}\right)\mathbb{P}_G+J\sum_{<ij>}(\hat{\mathbf{S}}_i.\hat{\mathbf{S}}_j-\frac{1}{4}\hat{n}_i\hat{n}_j),
\label{eq.tJModel}
\end{eqnarray}
which has been used extensively \cite{PALee2006} in the context of cuprates. Here the $a^\dagger_{i\sigma}$ is the usual fermion creation operator on site $i$, $n_i=\sum_\sigma
a^\dagger_{i\sigma}a_{i\sigma}$ and $\hat{\mathbf{S}}_i$ are the electron number and spin operators, respectively, and $\mathbb{P}_G=\prod_i(1-\hat{n}_{i\uparrow}\hat{n}_{i\downarrow})$ is a
projection operator excluding double occupancy of any site; $J$ can be estimated to be around $0.13$ eV \cite{PBourges1-2}. 

\section{Phase diagram and phenomena} \label{sec.cupratephenomena}
\begin{figure} [hbt]
\begin{center}
\includegraphics[width=13cm]{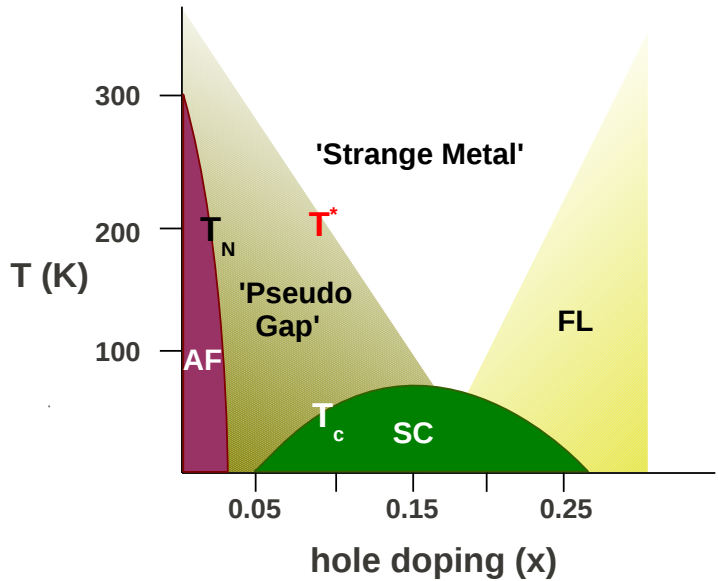}
\end{center}
\caption{\footnotesize{The famous phase diagram of hole doped cuprates, showing
superconductive (SC), antiferromagnetic (AF), \emph{pseudogap}, \emph{strange metal} and more conventional
Fermi liquid (FL) metal regions. Also indicated are the Ne'el temperature ($T_N$), superconducting transition
temperature ($T_c$) and somewhat \emph{ill defined} pseudogap temperature scale ($T^*$).}}
\label{fig.CupratePhaseDiagram}
\end{figure}

An idealized `phase diagram' of hole doped cuprates, in the hole density $x$ and temperature $T$ plane, is shown in Fig.\ref{fig.CupratePhaseDiagram}. Interestingly, it is nearly the same for the entire cuprate family. The parent compound is a
Mott insulator which undergoes a Ne'el transition at $T_N$ to an antiferromagnetic state. Upon doping, the $T_N$ drops very rapidly and goes to zero beyond a small value of doping ($x\simeq 0.03$). The
superconducting state sets in, preceded by a messy disordered low-temperature phase exhibiting spin glass like behavior till $x\simeq 0.05$. The superconducting transition temperature, $T_c(x)$, detected unambiguously by the onset of the zero resistance state, exhibits a well known parabolic (the `superconducting dome') dependence with increasing $x$ and reaches a maximum at $x\simeq 0.16$ and 
ultimately declines to zero for $x\apgt 0.25$. The $x$ value corresponding to the maximum $T_c$ is called the \emph{optimal doping} ($x_\mathrm{opt}$) and the regimes, $x<x_\mathrm{opt}$ and $x>x_\mathrm{opt}$, are denoted as
\emph{underdoped} and \emph{overdoped} regions, respectively. The high temperature region corresponding to the 
\emph{normal states} of the superconductor above the
superconducting dome is rather \emph{abnormal} ones among known states of matter found in Nature. This region can be broadly
classified into three parts, the so-called \emph{pseudogap} region for small $x$ (underdoped), the
\emph{strange metal} phase for intermediate values of $x$ and a more `normal' (i.e.~conventional) metal for
large $x$, i.e. the overdoped region. The pseudogap state goes into the strange metal phase, above a temperature scale depicted as the
pseudogap sclae $T^*$ (see Fig.\ref{fig.CupratePhaseDiagram}), which seems more like a crossover (as opposed to
a true phase transition) from variety of measurements. Below, we briefly discuss superconducting, strange metal phases and pseudogap phases and summarize some of their unusual features. We also include a separate detailed discussion of spectroscopic characteristics of these phases, as probed by angle-resolved photoemission (ARPES) and scanning tunneling (STS) spectroscopies.  

\subsection{Superconducting state}\label{Sec.SuperconductingState}

\begin{figure}[hbt]
\begin{center}
\begin{minipage}[c]{0.68\linewidth}
\includegraphics[width=\linewidth]{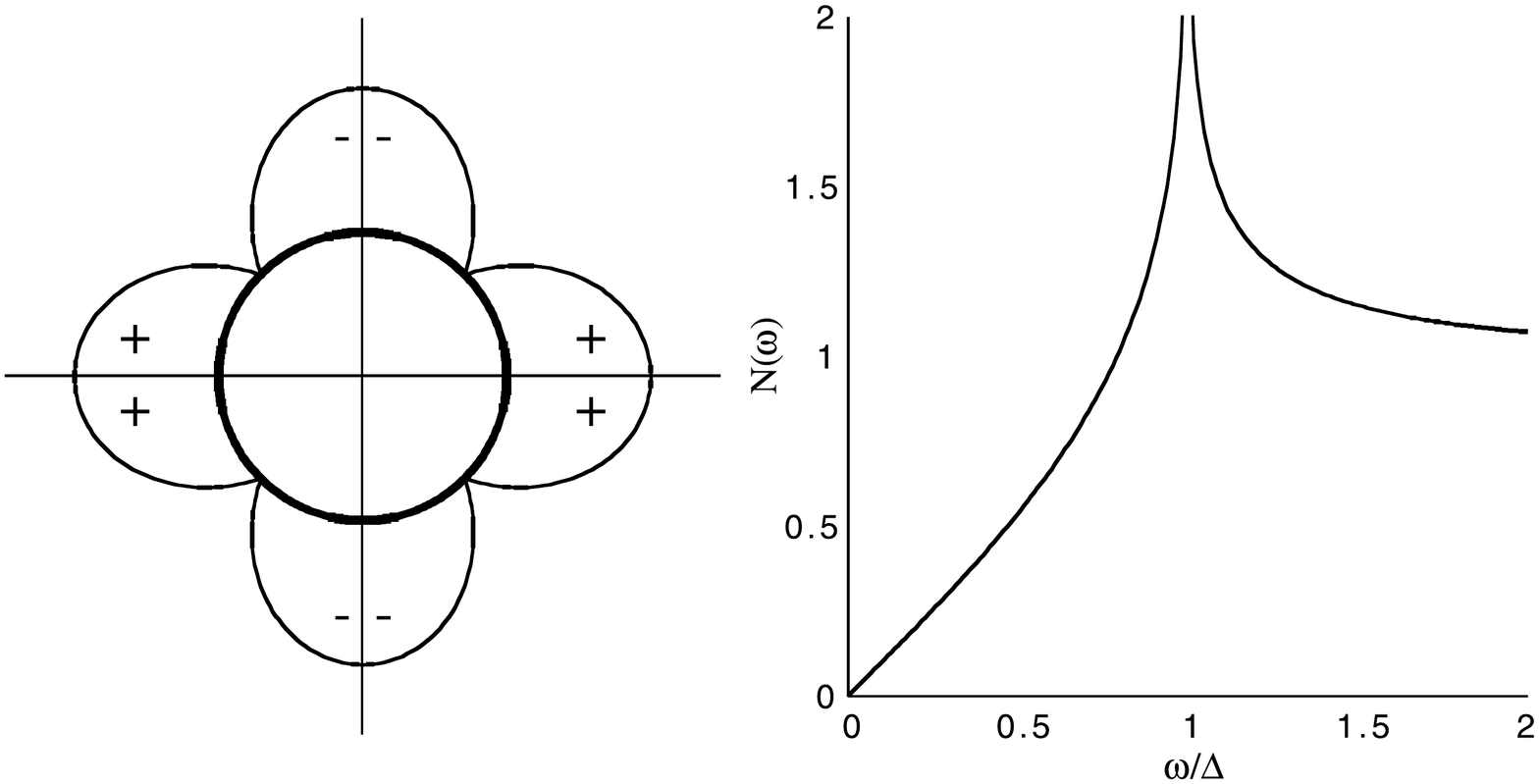}
\end{minipage}\hfill
\begin{minipage}[c]{0.28\linewidth}
\caption{{\footnotesize Variation of the $d$-wave gap on the Fermi surface is shown in the left figure, $k_x=\pm k_y$ is the nodal direction. The density of states (DOS) is plotted in the right figure. A DOS divergence appears at the maximum gap magnitude $\omega=\Delta$. (From Ref.\onlinecite{MRNorman1})}\label{fig.dWaveGap}}
\end{minipage}
\end{center}
\end{figure}

The superconducting state is less unconventional compared to the other parts of high-$T_c$ phase diagram. The
superconducting order parameter is a charge $2e$ complex field with $d$-wave symmetry as has been unambiguously demonstrated by phase
sensitive measurements \cite{DAWollman1-2,CCTsuei1-2}. 
In conventional superconductors the Cooper pair amplitude or the energy gap
\cite{MTinkham1} that appears in the
spectrum of electronic excitations with momentum $\mathbf{k}$ on the Fermi surface, namely 
$\Delta_\mathbf{k}$, is generally independent of $\mathbf{k}$,
whereas in cuprates it is strongly $\mathbf{k}$ dependent and has nodes i.e.~vanishes along certain
$\mathbf{k}$ directions. An approximate two dimensional representation of $d$-wave symmetry superconducting
gap, whose validity for cuprates has been established by ARPES \cite{ZXShen1,ADamascelli1,JCCampuzano1} and phase-sensitive Josephson junction measurements (see later), is 
$\Delta_\mathbf{k}=(\Delta_0/2)(\cos{k_xa}-\cos{k_ya})$ (see Fig.\ref{fig.dWaveGap}, $a$ is Cu-O-Cu bond length in the plane and $\Delta_0$ is the maximum gap magnitude). This can arise in a lattice system through nearest neighbor
pairing such that magnitude of pair amplitude (a complex number) along $x$ and $y$ bonds is same but the
phases differ by $\pi$ (see Fig.\ref{fig.YBCO_Crystal}). A gap function of this kind implies nodal Bogoliubov quasiparticles leading to substantial density of gapless excitations (see Fig.\ref{fig.dWaveGap}) and these would dominate the behavior of various physical properties at low temperatures ($T\ll T_c$). Particularly, since the density of states (DOS) vanish linearly in
energy, the nodal excitations give rise to, e.g., a non-activated $T^2$ term in electronic specific heat\cite{BNMomono1-2} and similar power law $T$ dependences of various other quantities \cite{JAMartindale1-2,WNHardy1-2}, unlike that in 
a $s$-wave superconductor. An interesting effect \cite{GVolovik1-2} can be seem in the presence of a magnetic field $H$ due to a shift of the quasiparticle spectrum by an amount $\propto H^{1/2}$; this has been verified experimentally \cite{KAMoller1-2}.

 The superconductivity in cuprates has several unusual characteristics. Here we list a few of them. 

{\bf 1. Short coherence length:} The length scale associated with pairing, namely the coherence length ($\xi_0$), is very small in cuprate superconductors - around $15-20~\AA$ in the plane and, between planes $2~\AA$, even smaller than the interlayer spacing, implying that the superconductivity related phenomena in the cuprates can be thought of as effectively having Josephson coupled \cite{MTinkham1} planes.

\begin{figure}[hbt]
\begin{center}
\begin{minipage}[c]{0.48\linewidth}
\includegraphics[width=\linewidth]{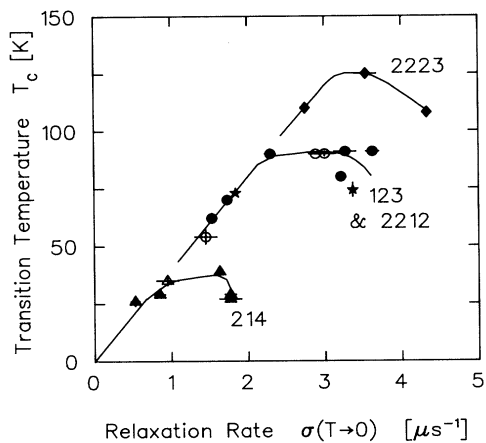}
\end{minipage}\hfill
\begin{minipage}[c]{0.48\linewidth}
\caption{{\footnotesize The superconducting transition temperature $T_c$ as a function of low
temperature muon spin-relaxation rate $\sigma(T\rightarrow0)\propto 1/\lambda_{ab}^2(0)\propto\rho_s(0)$ for
different cuprate compounds. At the low doping values data fall on the
\emph{universal} straight line indicating the Uemura correlation $\rho_s\propto T_c$. (See Ref.\onlinecite{YJUemura1989} for more details).}\label{fig.UemuraScaling}}
\end{minipage}
\end{center}
\end{figure}

 {\bf 2. Low superfluid density:} Cuprates superconductors are characterized by very small in-plane superfluid density or stiffness $\rho_s$. The
low temperature superfluid density $\rho_s$ for small doping goes as $x$, as also does $T_c$. This is the so-called Uemura scaling
\cite{YJUemura1989}, $\rho_s\propto T_c$ for small $x$ (see
Fig.\ref{fig.UemuraScaling}). Such a scaling is unusual for conventional superconductors, typically having very large $\rho_s$ and much smaller
$T_c$ \cite{VJEmery2-2}. Since the in-plane penetration depth $\lambda_{ab}\propto 1/\rho_s^{1/2}$,
cuprates have very large penetration depth ($\sim 1500~\AA$), implying the extreme type-II limit \cite{MTinkham1} for these
superconductors. Small superfluid density of phase stiffness indicates the importance of phase fluctuations, which are expected
to be very strong, especially in the underdoped regime. This fact is corroborated by the observation of large Nernst effect\cite{YWang1-2} and enhanced fluctuation diamagnetism\cite{LLi1-2} by Ong and coworkers over an unusually large temperature range above $T_c$. A phase fluctuation related theory for this is described later here. 

{\bf 3. Gap to $T_c$ ratio:} Another important experimental observation about the superconducting state is that 
the zero temperature superconducting gap 
$\Delta_0$ inferred, e.g., from STM \cite{OFischer1} or ARPES \cite{ADamascelli1,JCCampuzano1} increases, whereas 
$T_c$ declines, as mentioned above, in the underdoped side with decreasing doping. Both the measured ratio $(2\Delta_0/k_\mathrm{B}T_c)$ and the $x$ dependence imply the
violation of the BCS relation between $\Delta_0$ and $T_c$ i.e.~ $2\Delta_0/k_BT_c\simeq 4.3$ for $d$-wave\cite{HWon1}. However, of late, the large gap seen via spectroscopy has been attributed to other competing order \cite{LTaillefer2009,PALee2014} or as having non-pairing origin \cite{Kyung2006,Ferrero2009,WWu2017}.

{\bf 4. Emergence from pseudogap state:} In the underdoped side, the precedence
of \emph{pseudogap phase} (see Fig.\ref{fig.CupratePhaseDiagram}) while approaching the superconducting state
from high temperatures suggests that the transition is not a sudden gap appearance one, as in BCS. This is related with the pseudogap phenomena which we discuss below in more detail. 

{\bf 5. Anomalous high-field vortex states:} The mixed state of cuprate superconductor at high field and low temperature is rather anomalous, showing mangetoconductivity (Shubnikov-deHaas (SdHvA)) and magnetization (deHaas-van Alphen (dHvA)) oscillations as a function of $(1/H)$, e.g.~in clean $\mathrm{YBa_2Cu_4O_8}$ and $\mathrm{HgBa_2CuO_{4+\delta}}$ \cite{Doiron2007,SSebastian2008,NBarisic2013}. These quantum oscillations have been seen with an accompanying long-range charge-density wave (CDW) order even far below the irreversiblity field for superconducting to non-superconducting transition \cite{TWu2011,DLeBoeuf2013,SGerber2015}, presumably related to vortex lattice or vortex glass depinning \cite{FYu2016}.

\subsection{Strange metal} \label{Sec.StrangeMetal}
 
 The strange metal is one of the most enigmatic phases in the cuprate phase diagram (Fig.\ref{fig.CupratePhaseDiagram}). Transport measurements reveal that the resistivity is linear in temperature over a large range, the most fascinating example being the single layer Bi2201, where linearity persists down to $T_c\simeq 7$ K \cite{SMartin1-2}, starting from an amazingly high temperature 
($\sim 700$ K).  This behaviour is in strong
contrast to what one expects on the basis of Fermi liquid theory. We discuss a few of the strange properties of the strange metal phase below.  

\begin{figure}[hbt]
\begin{center}
\begin{minipage}[c]{0.48\linewidth}
\includegraphics[width=\linewidth]{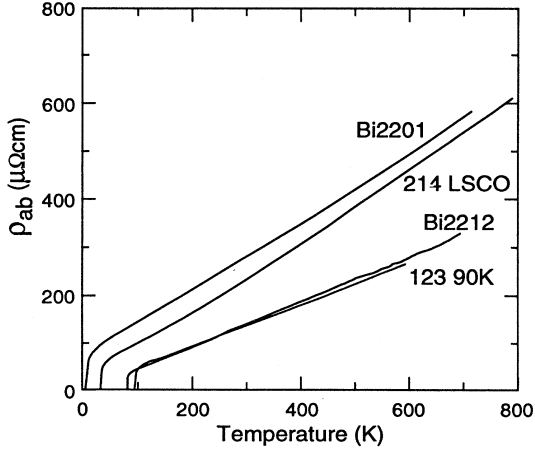}
\end{minipage}\hfill
\begin{minipage}[c]{0.38\linewidth}
\caption{{\footnotesize Temperature dependence of the in-plane resistivity $\rho_{ab}$ for various optimally doped cuprates. (From
Ref.\onlinecite{BBatlogg1-2}).} \label{fig.Resistivity_Cuprates}}
\end{minipage}
\end{center}
\end{figure}

{\bf 1. $T$-linear resistivity:} In conventional low-temperature superconductors, the temperature dependence of resistivity $\rho$ is well described by $\rho\approx \rho_0 + b T^5$, at low $T$ ($>T_c$). This $T$ dependence well below the phonon Debye temperature $\theta_\mathrm{D}$ arises from the scattering of
electrons by phonons. At higher temperatures ($T\apgt \theta_\mathrm{D}$), a linear behavior is expected, and the interpolation between
the two regimes is given by the Gr$\mathrm{\ddot{u}}$neisen-Bloch formula. The residual
resistivity $\rho_0$ at $T=0$ is caused by scattering by impurities. However, the behaviour of $\rho$
observed in high-$T_c$ superconductors is strikingly different. Resistivity in the
$\mathrm{CuO_2}$ planes for several different cuprate compounds near optimal doping exhibits an approximate linear behavior
over a large temperature window \cite{BBatlogg1-2,HTakagi1}  as shown in Fig.\ref{fig.Resistivity_Cuprates}. The resistivity values are very large at high temperature, e.g., in comparison to that of copper at room temperature, which has a resistivity of $1.7~\mu\Omega\mathrm{cm}$. This seems to be one of the defining characteristics of strongly correlated metals i.e.~typical resistivity values are quite large, often of the same order or even larger than the Ioffe-Regel limit and increase linearly with temperature without any saturation. Fig.\ref{fig.Resistivity_Cuprates} demonstrates clear linear behaviour with very similar slopes for the various compounds, suggesting a possible common origin for in-plane transport.

 The linear $T$ dependence \cite{HTakagi1} is something not fundamentally understood, and is generally associated with
strong correlation in the system. As is well known, electron-electron interaction in Fermi liquid systems leads to a
$T^2$ dependence of resistivity, unlike what is observed in optimally doped cuprates. With increasing doping,
the temperature dependence is of the form $\rho_{ab}\propto T^\alpha$ with $\alpha$ increasing smoothly from 1
to 2 from optimum to overdoped \cite{NEHussey1}. This behaviour is also not understood, but presumably can be broadly 
described as due to the system changing from strongly correlated non-Fermi liquid to weakly interacting Fermi liquid.

\begin{figure}[hbt]
\begin{center}
\includegraphics[height=7cm]{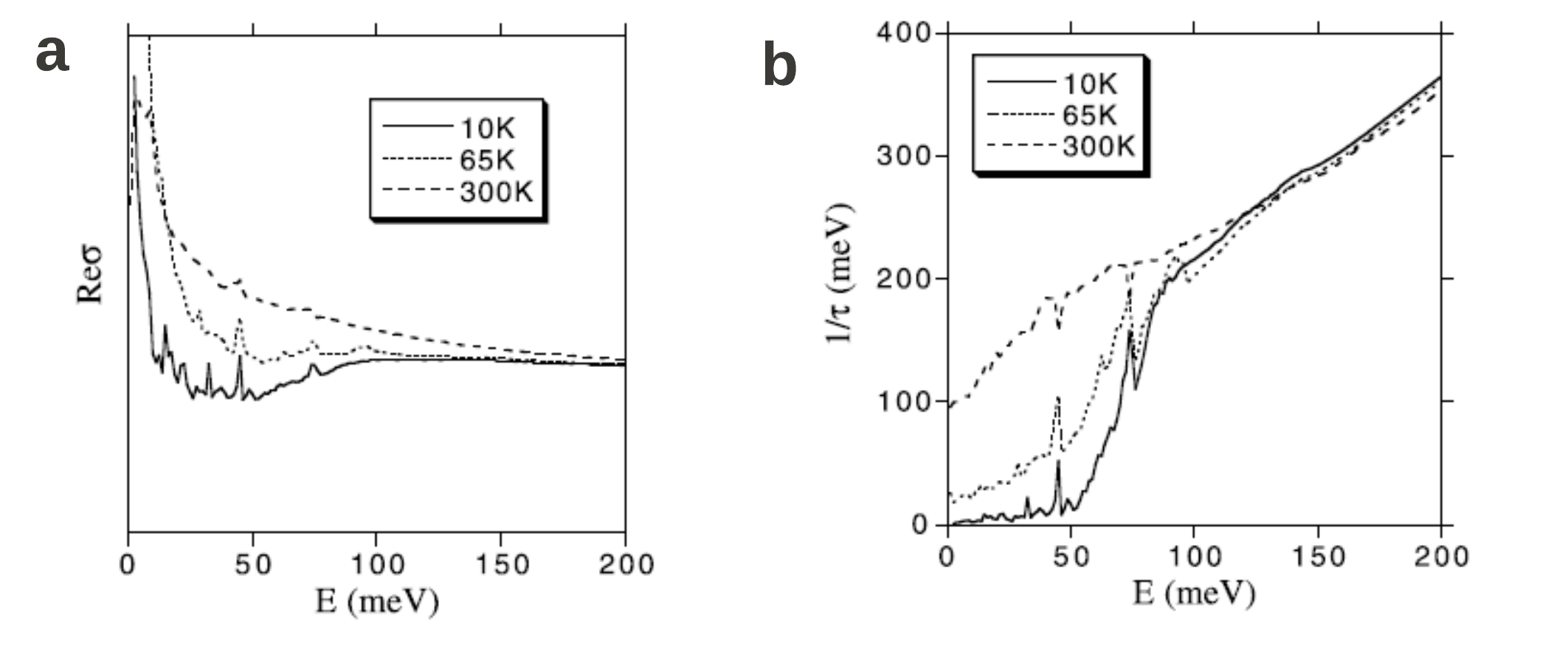} 
\end{center}
\caption{\footnotesize{ {\bf a}, Real part of in-plane infrared optical conductivity of underdoped YBCO for the superconducting state (10 K), pseudogap phase (65 K), and normal state
(300 K). A gap-like suppression develops at around 50 meV in the pseudogap regime. {\bf b}, Variation of extracted optical scattering rate $1/\tau(\omega)$ with frequency for underdoped YBCO. (From Ref.\onlinecite{AVPuchkov1}).}
\label{fig.OpticalConductivity}}
\end{figure}

{\bf 2. $\omega$-linear scattering rate:} Optical conductivity, which probes the current-current correlation function, gives further evidence of the unconventional nature of the strange metal phase. 
In plane, the normal state (both strange metal and pseudogap states) of cuprates is characterized
by a broad, Drude-like response centered at frequency $\omega=0$ (Fig.\ref{fig.OpticalConductivity} {\bf a}) for optical conductivity $\sigma(\omega)$. The data are phenomenologically represented by a generalized Drude form \cite{AVPuchkov1}
\begin{eqnarray}
\sigma(\omega)&=&\frac{1}{4\pi}\frac{\omega^2_\mathrm{pl}}{1/\tau(\omega)-i\omega[1+\lambda(\omega)]},
\label{eq.GeneralizedDrude}
\end{eqnarray}   
where $\omega_\mathrm{pl}$ (the plasma frequency) is given by the sum rule $\int_0^\infty d\omega
\mathrm{Re}\sigma(\omega)=(\omega^2_\mathrm{pl}/8)$ (here, $\sigma=\mathrm{Re}\sigma+i\mathrm{Im}\sigma$ and the integral has a high frequency cut off $\sim 1$ eV). In the above form, $1/\tau(\omega)$ and $1+\lambda(\omega)$ are the frequency dependent scattering rate and mass renormalization, respectively. The analysis of the experimental data using eq.\eqref{eq.GeneralizedDrude} reveals that the scattering
rate has the form $a+b\omega$ (see Fig.\ref{fig.OpticalConductivity} {\bf b}), unlike a Fermi liquid where $1/\tau\sim \omega^2$. The $\omega$-linear term is expected from a
marginal Fermi liquid theory \cite{EAbrahams1-2,CMVarma2-2}.  

\subsection{Pseudogap \emph{phase}} \label{Sec.PseudogapPhase}

  One of the most controversial aspects of the cuprate phase diagram is the nature of \emph{pseudogap phase}.
This is associated with a set of ill-understood apparent crossover phenomena \cite{TTimusk1,SHufner1,MRNorman2,JLTallon1}, 
which are widely observed in underdoped cuprates and, to some extent, in optimally and even slightly overdoped
materials. As we summarize below, signatures of gap are observed through different experimental probes, e.g., spin susceptibility, thermodynamic, transport and optical measurements, and in the quasiparticle excitations
investigated through ARPES and scanning tunneling microscopy (STM),. As shown in
Fig.\ref{fig.CupratePhaseDiagram}, these gap-like features (`pseudogap' as opposed to true gap) become prominent below a 
temperature scale $T^*$ (the pseudogap temperature), which varies somewhat when detected using different
probes. In the underdoped samples, $T^*$ is way above the superconducting transition temperature $T_c$, which
is traditionally detected through the onset of zero resistance state, indicating that superconducting long range order (LRO)
is not present in the pseudogap state even though there are clear hallmarks of a gap of the kind resembling that of the
$d$-wave symmetry superconducting state. Below, we touch upon  some of the salient fetures of the pseudogap phase.

\begin{figure}[hbt]
\begin{center}
\begin{minipage}[c]{0.48\linewidth}
\includegraphics[width=\linewidth]{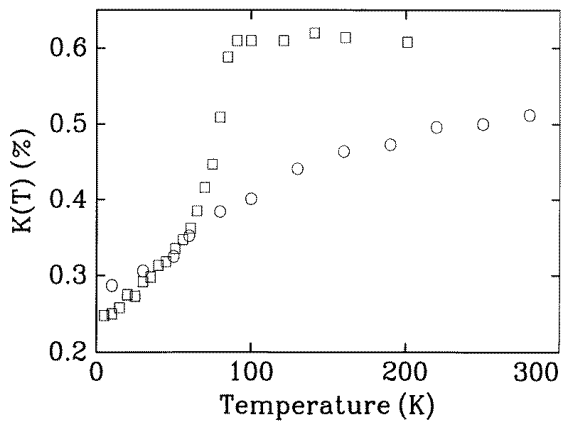}
\end{minipage}\hfill
\begin{minipage}[c]{0.38\linewidth}
\caption{{\footnotesize  Knight shift in $\mathrm{YBa_2Cu_3O_{6.95}}$ (squares) and
underdoped $\mathrm{YBa_2Cu_3O_{6.64}}$ (circles). The normal-state susceptibility is temperature independent above $T_c$ in the optimally doped compound, but decreases
with temperature in the underdoped compound even above $T_c$. (From Ref.\onlinecite{TTimusk1}).} 
\label{fig.KnightShift}}
\end{minipage}
\end{center}
\end{figure}

{\bf 1. ``Spin gap" in spin susceptibility:} The first experimental indication \cite{WWWarren1} of a pseudogap phase came from NMR measurements in underdoped YBCO, which showed
that the spin lattice relaxation rate (which depends on the number of conduction electron spin states at the Fermi energy) starts to decrease much above $T_c$. A similar decrease is seen in the related Knight shift \cite{HAlloul1} (Fig.\ref{fig.KnightShift}). NMR
experiments probe the spin channel, unlike optical conductivity which probes the charge degrees of freedom.
The decrease of Knight shift indicates decrease of spin susceptibility with temperature in the pseudogap state of cuprates, initially supporting the hypothesis of spin gap \cite{WWWarren1}. Subsequent experiments, however, have demonstrated that the pseudogap exists in both spin and charge channels, as evident from, e.g., ARPES or optical
conductivity. 

\begin{figure}[hbt!]
\begin{center}
\begin{minipage}[c]{0.45\linewidth}
\includegraphics[width=\linewidth]{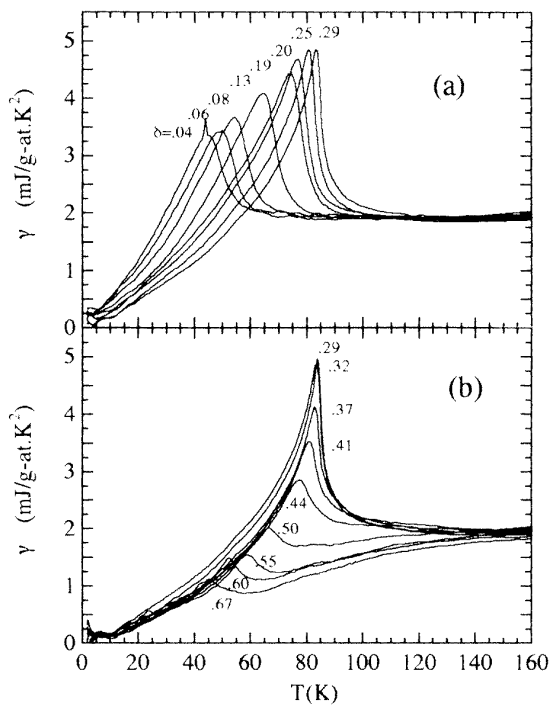}
\end{minipage}\hfill
\begin{minipage}[c]{0.38\linewidth}
\caption{{\footnotesize Specific heat coefficient $\gamma$ for (a) overdoped and (b) underdoped
$\mathrm{Y_{0.8}Ca_{0.2}Ba_2Cu_3O_{7-\delta}}$. In the overdoped material a gap opening, signalled by a peak and the subsequent suppression of $\gamma$, occurs below $T_c$. In the underdoped materials a gap appears to form much above $T_c$, around 140K, in the normal state itself. (From Ref.\onlinecite{JWLoram7}).}\label{fig.SpecificHeat}}
\end{minipage}
\end{center}
\end{figure}

{\bf 2. Suppression of specific heat:} Another evidence of pseudogap comes from the electronic specific heat \cite{JLTallon1,TTimusk1}. The electronic
specific heat provides thermodynamic evidence for a gap in the normal state of the cuprates \cite{JWLoram1}. The specific heat is generally found to be linear in temperature $C_v\approx \gamma T$ above the pseudogap temperature scale $T^*$. Below $T^*$, $C_v/T$ begins to decrease with decreasing temperature
(Fig.\ref{fig.SpecificHeat}), indicating presence of a normal state gap. For YBCO, above $T_c$, overdoped samples 
show (Fig.\ref{fig.SpecificHeat} {\bf a}) a temperature- and doping-independent $\gamma$. The underdoped samples have a depression of the specific heat coefficient in the normal state below $T^*$ 
(Fig.\ref{fig.SpecificHeat} {\bf b}) \cite{JWLoram1} and the specific heat jump at $T_c$ is diminished with decreasing doping. Evidence of pseudogap in specific heat measurements has also been found in
other cuprates \cite{JWLoram2,JWLoram3}.

\begin{figure}[hbt]
\begin{center}
\includegraphics[height=6cm]{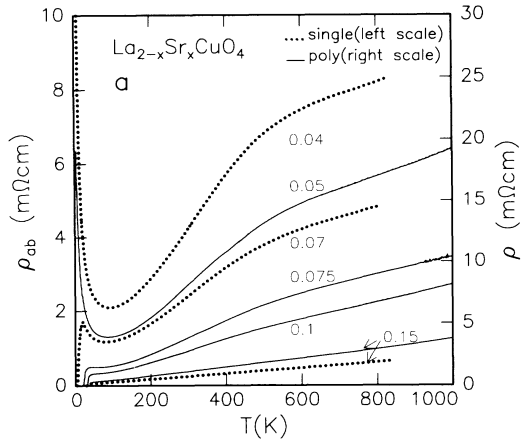}  
\end{center}
\caption{\footnotesize{Temperature dependence of in-plane resistivity for underdoped LSCO ($0\leq x\leq 0.15$). Around the pseudogap temperature scale, the resistivity starts deviating from nearly linear in $T$ high-temperature behavior. (From Ref.\onlinecite{HTakagi1}).}
\label{fig.Resistivity_LSCO}}
\end{figure}

{\bf 3. Transport signatures of pseudogap:} There is a significant deviation of in-plane or the $ab$-plane resistivity from the $T$ linear temperature dependence of strange metal at higher temperature. A pseudogap temperature \cite{HTakagi1} is identified as the point below which temperature-dependence of in-plane resistivity deviates significantly from its high temperature
behavior (see, e.g., Fig.\ref{fig.Resistivity_LSCO} {\bf a}). A similar temperature scale can be inferred from 
the Hall resistance which, quite remarkably, is strongly temperature dependent \cite{HYHwang1}. The pseudogap also appears in the $c$-axis resistivity $\rho_c$ \cite{KTakenaka1,ANLavrov1} as an insulator-like increase with decreasing temperature.

\begin{figure}[hbt!]
\begin{center}
\begin{minipage}[c]{0.48\linewidth}
\includegraphics[width=\linewidth]{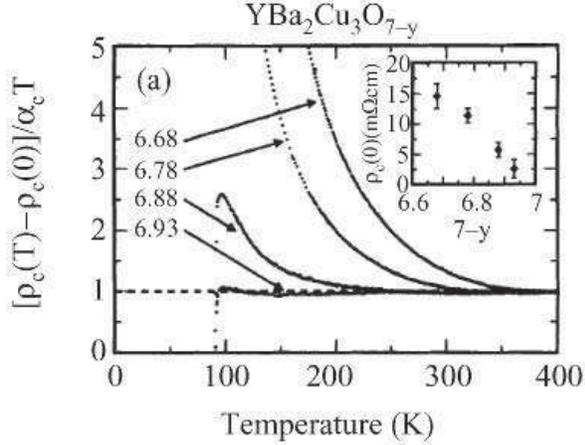}
\end{minipage}\hfill
\begin{minipage}[c]{0.38\linewidth}
\caption{{\footnotesize The temperature dependence of the $c$-axis resistivity in underdoped and optimally
doped $\mathrm{YBa_2Cu_3O_{7-\delta}}$ after the subtraction of the intercept of a approximately linear-$T$ metallic part; $\alpha_c$ is the slope of the linear-$T$ part. (From Ref.\onlinecite{KTakenaka1}).}\label{fig.cAxisResistivity}}
\end{minipage}
\end{center}
\end{figure}

{\bf 4. Optical gap:} A signature of a gap is also observed in infrared measurements exhibiting a suppression in the $ab$-plane optical conductivity $\sigma'_{ab}(\omega)$ separating the low energy Drude peak from `bump'-like feature at mid-infrared
\cite{SLCooper1,LDRotter1} (see Fig.\ref{fig.OpticalConductivity} {\bf a}). Interestingly, $c$-axis optical conductivity $\sigma_c(\omega)$ reveals more dramatic effect of a pseudogap; $\sigma_c(\omega)$ does not have a Drude peak at low energies and shows a significant gap below a characteristic frequency scale. The gap becomes more prominent as one approaches $T_c$.

\begin{figure}[hbt!]
\begin{center}
\begin{minipage}[c]{0.48\linewidth}
\includegraphics[width=\linewidth]{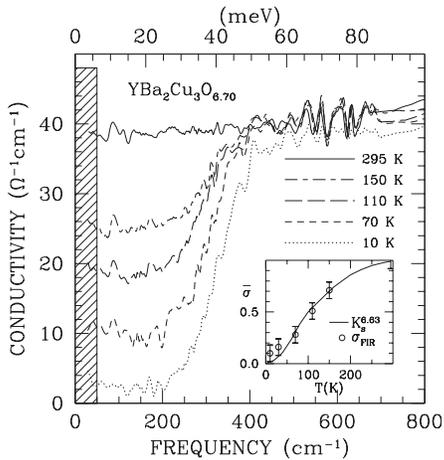}
\end{minipage}\hfill
\begin{minipage}[c]{0.38\linewidth}
\caption{{\footnotesize $c$-axis optical conductivity for underdoped YBCO demonstrating a pseudogap which fills in with
temperature (From Ref.\onlinecite{AVPuchkov1})}\label{fig.cAxisOpticalConductivity}}
\end{minipage}
\end{center}
\end{figure}

{\bf 5. Temperature and doping dependent Drude peak:} The other interesting observation is that
the width of the Drude peak in $\sigma_{ab}$ decreases with temperature, but the area (spectral weight) under
it is independent of temperature \cite{AFSantander1}. Intriguingly, the integrated area under the Drude peak is found to 
be linear in $x$ \cite{JOrenstein1,SUchida1} indicating a connection to the doped holes \cite{PALee2006} and this
weight transforms to form the delta function superfluid peak in the superconducting state suggesting $\rho_s\propto x$. However, as we discuss below, ARPES shows a Fermi surface with an area corresponding to $1-x$ electrons \cite{PALee2006}.

\begin{figure}[hbt!]
\begin{center}
\includegraphics[height=9cm]{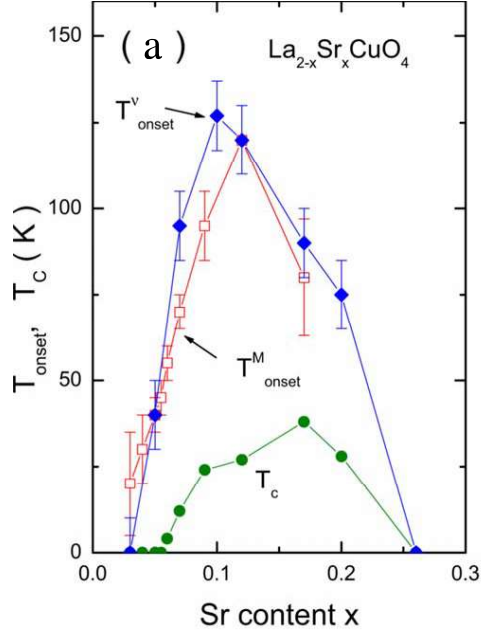}
\end{center}
\caption{\footnotesize The phase diagram of LSCO showing the onset temperatures, $T^{\nu}_\mathrm{onset}$ and $T^M_\mathrm{onset}$, determined by enhanced Nernst and diamagnetic signal, respectively, as a function of doping. $T_c(x)$ is shown for comparison (From Ref.\onlinecite{Li2010}).}
\label{fig.NernstOnset}
\end{figure}

{\bf 6. Anomalous Nernst and fluctuation diamagnetism regime in the pseudogap phase:}
Experiments have found a very large diamagnetic and Nernst response in the pseudogap phase of the cuprates. A large diamagnetic signal naturally points towards fluctuating superconductivity as one of the possible origins. Also, the fact that the Nernst response is usually very small in typical nonmagnetic metals and a much stronger response is observed in the vortex-liquid regime of superconductors supports this point of view. Compared to conventional superconductors, the Nernst and diamagnetic response have been found to exist over an anomalously large region \cite{Wang2006,Li2010} in the pseudogap phase, extending to temperatures far above the superconducting transition temperature $T_\mathrm{c}$ (see however ref.~\onlinecite{Chang2012,Yu2012,Kokanovic2014}). Another important piece of the debate is related to the fact that the putative boundary \cite{Wang2006,Li2010} of the large Nernst and diamagnetic response regime, the so-called `onset' temperature $T_\mathrm{onset}$, tracks $T_\mathrm{c}(x)$ and so follows a dome-shaped curve as a function of doping $x$ instead of tracking the pseudogap temperature scale $T^*(x)$, which monotonically decreases with $x$ (see Fig.\ref{fig.NernstOnset}) . This has been argued as evidence against a pairing origin of the pseudogap, mainly due to the expectation that if the pseudogap arises from pairing then superconducting fluctuations and associated Nernst and diamagnetic responses should persist all the way up to pseudogap temperature~\cite{PALee2006}. 

\subsection{Spectroscopic characteristics of superconducting, strange metal and pseudogap phases} \label{Sec.Spectroscopy}

Angle-resolved photoemission spectroscopy (ARPES) and scanning tunneling microscopy (STM) have emerged as two of the major tools for probing condensed matter systems mainly due to their success in directly accessing electronic excitations in cuprates. Here we discuss in somewhat more detail the major findings and phenomenologies that have emerged out of the  ARPES and STM studies on cuprates. 

\begin{figure}[hbt!]
\begin{center}
\includegraphics[height=9cm]{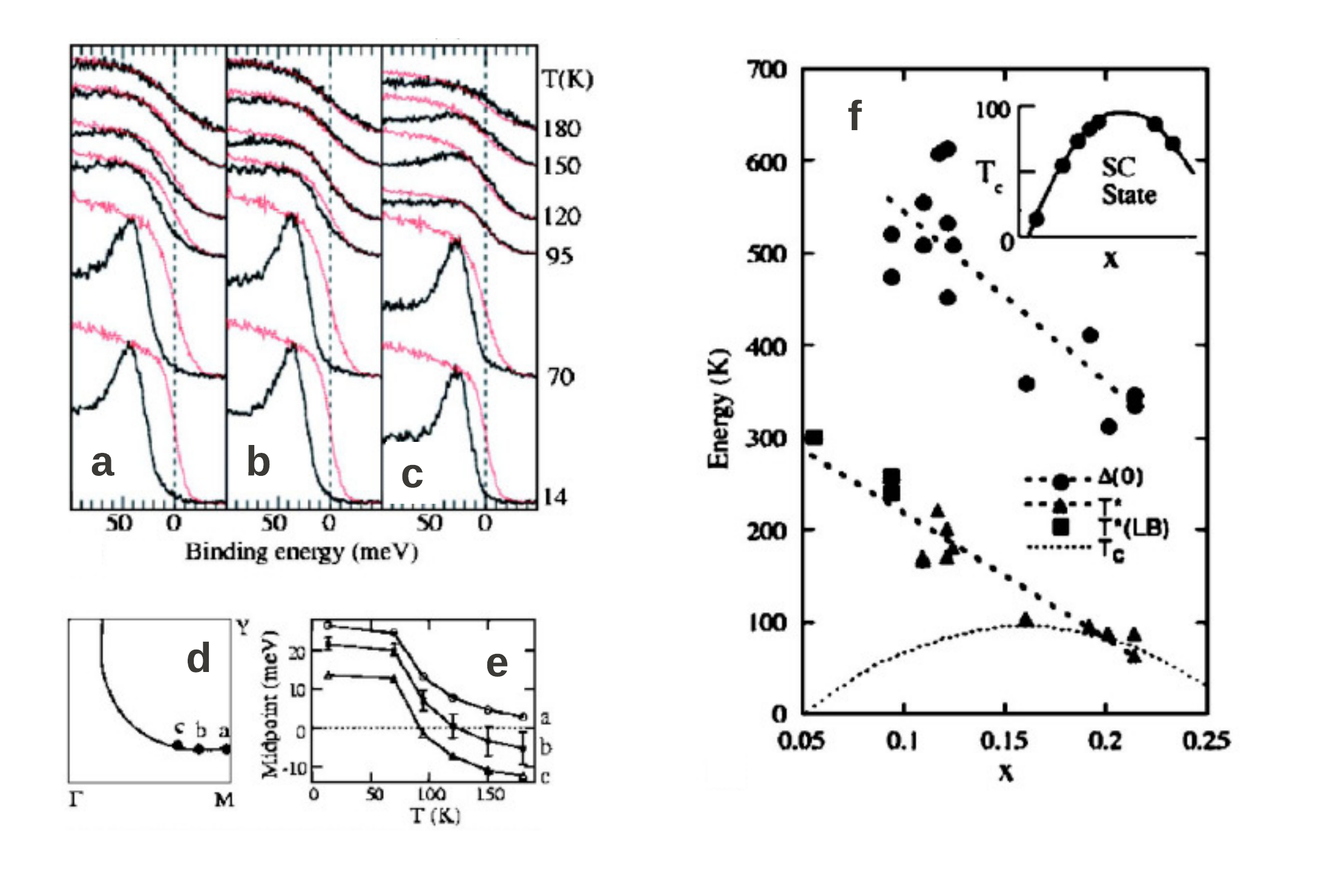} 
\end{center}
\caption{{\footnotesize ARPES of underdoped cuprates. {\bf a}-{\bf c} Unsymmetrized energy distribution curves (EDC) for underdoped 
Bi2212 $T_c=85$ K at different $\mathbf{k}$ points on the Fermi surface (panel {\bf d}). The existence of a gap for $T>T_c$ is signaled by the pullback of the spectrum from the Fermi surface determined by the Pt reference (red lines). {\bf e}, Temperature 
dependence of the gap determined from the leading-edge midpoints. (From Ref.\onlinecite{MRNorman3}). {\bf f}, The temperature $T^*$ corresponding to the appearance of the pseudogap is plotted as a function of doping for Bi2212. Triangles
are determined from data shown in {\bf a}. Circles show the
energy gap $\Delta (0)$ measured at the antinodal point at low temperatures in the superconducting state. (From Ref.\onlinecite{JCCampuzano1}).}}
\label{fig.EDC}
\end{figure}

\begin{figure}[hbt]
\begin{center}
\includegraphics[height=9cm]{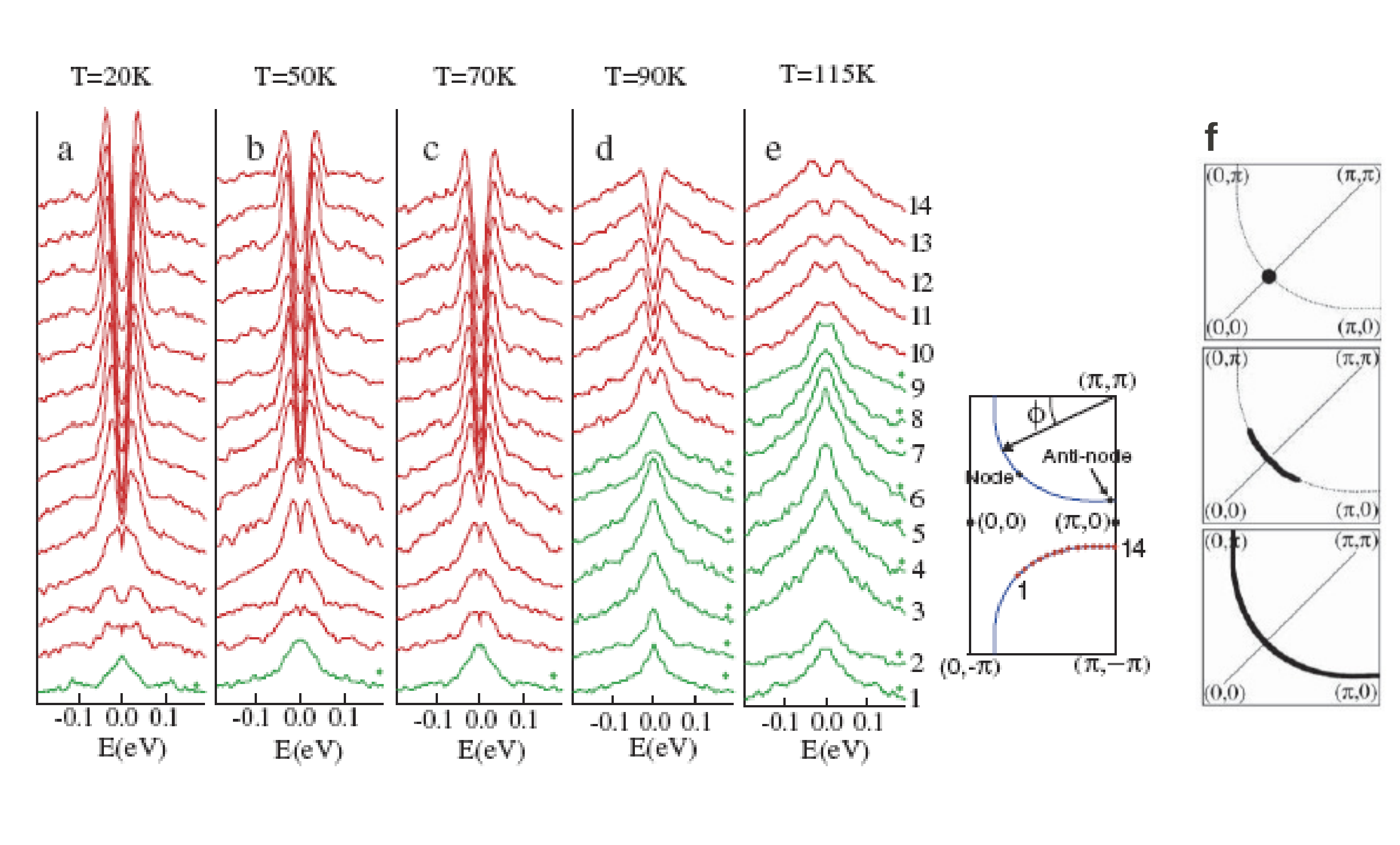} 
\end{center}
\caption{{\footnotesize {\bf a-e}, Evolution of the symmetrized energy distribution curves (EDCs) for underdoped Bi2212 $Tc=80$ K with temperature. The different EDCs correspond to different $\mathbf{k}$ points from the node (lowest curve) to the 
antinode (uppermost curve). Gapless spectra having a peak at zero energy are shown in green. (From Ref.\onlinecite{AKanigel1}). {\bf f}, Schematic illustration
of the development of Fermi arcs with temperature in the pseudogap phase. The $d$-wave node below $T_c$ (top panel) 
becomes an extended gapless arc above $T_c$ (middle panel) which expands in length with increasing $T$ to form the full Fermi
surface at $T^*$ (bottom panel). (From Ref.\onlinecite{JCCampuzano1}).}}
\label{fig.SymmetrizedEDC}
\end{figure}

{\bf 1. ARPES:} It is really the ARPES experiments \cite{ADamascelli1,JCCampuzano1} that unambiguously demonstrated the pseudogap. Reviews by Damascelli {\it et al.} \cite{ADamascelli1} and 
Campuzano {\it et al} \cite{JCCampuzano1} provide an in-depth discussion on this topic. Cuprates are ideal candidates for ARPES experiments owing to their approximate 2d nature that enables easy cleavage and possibility of experiments on atomically 2d surfaces that are representative of the bulk. One can also infer the in-plane electron momentum $\mathbf{k}$ with high precision. ARPES, a modern version of 
familiar photoelectric effect, measures the spectrum of electrons (photoemission intensity as a function of 
energy $\omega$ and momentum $\mathbf{k}$ of electron in the sample) ejected due to incident photons. In this way, ARPES 
probes the electronic spectral function or spectral density $A(\mathbf{k},\omega)$, that quantifies the probability
of finding an electron with momentum $\mathbf{k}$ and energy $\omega$ (measured from the Fermi energy or the
chemical potential). The energy corresponding to the
position of a peak (if there is one, as in the case of a quasiparticle in a Fermi liquid) in 
$A(\mathbf{k},\omega)$ for a fixed $\mathbf{k}$ gives the information of energy of electronic excitations for that $\mathbf{k}$, and their dispersion; the width of the peak is a measure of the life time of such excitations. ARPES can map out the
Fermi surface from the $\mathbf{k}$-space locus of the minimum energy excitation \cite{ADamascelli1,JCCampuzano1} in the 
Brillouin zone. Peak in $A(\mathbf{k},\omega)$ for a particular $\mathbf{k}$ on the Fermi
surface appearing at a nonzero energy indicates a gap in the electronic excitation. The gap can either be
traced from the raw data of energy distribution curve (EDC, photoemission intensity as a function of energy for a fixed
$\mathbf{k}$) through the pulling back of the leading edge of the electronic spectrum from the Fermi energy 
(Figs.\ref{fig.EDC} {\bf a}-{\bf c}) or from the peak position of symmetrized EDC \cite{ADamascelli1,JCCampuzano1}
(Figs.\ref{fig.SymmetrizedEDC} {\bf a}-{\bf e}).

 Angle-resolved photoemission spectroscopy shows that an energy gap is observed on the Fermi surface crossing
along the direction $(\pi,0)$ to $(\pi,\pi)$ i.e.~at the antinodal point, even above $T_c$ for underdoped
sample (Figs.\ref{fig.EDC} {\bf a}-{\bf c}). Although, above $T_c$, a finite spectral weight can be detected
at zero energy ($\omega=0$), justifying the nomenclature `pseudogap'. With increasing
temperature the pseudogap \emph{fills in} rather than closing in. Intriguingly, the spectral line width is large implying an incoherent spectra above $T_c$. The appearance of a coherent 
peak at the gap edge (Figs.\ref{fig.EDC} {\bf a}-{\bf c}) indicates the appearance of superconductivity. The size of the pullback of the leading edge and the energy gap measured by the location of the coherence peak in the superconducting state are basically same. As shown in Fig.\ref{fig.EDC} {\bf f}, the estimated gap increases with decreasing doping in contrast to $T_c$ which decreases.

 As shown in Fig.\ref{fig.SymmetrizedEDC}, below $T_c$, the energy gap is maximal near $(0,\pi)$ and vanishes along the line connecting $(0,0)$ and $(\pi,\pi)$
i.e. the nodal direction ($k_x=k_y$), giving rise to gapless nodal quasiparticles, confirming the $d$-wave nature of the superconductivity. Above
$T_c$ the gapless region expands to cover a finite region near the nodal point giving rise to the so-called `Fermi arcs'\cite{HDing1,ACLoeser1,DSMarshall1} (see Fig.\ref{fig.SymmetrizedEDC} {\bf f}). Strangely, spectral line shape is relatively sharp in the nodal direction even above $T_c$, unlike that along the antinodal direction. 

\begin{figure}[hbt]
\begin{center}
\begin{minipage}[c]{0.68\linewidth}
\includegraphics[width=\linewidth]{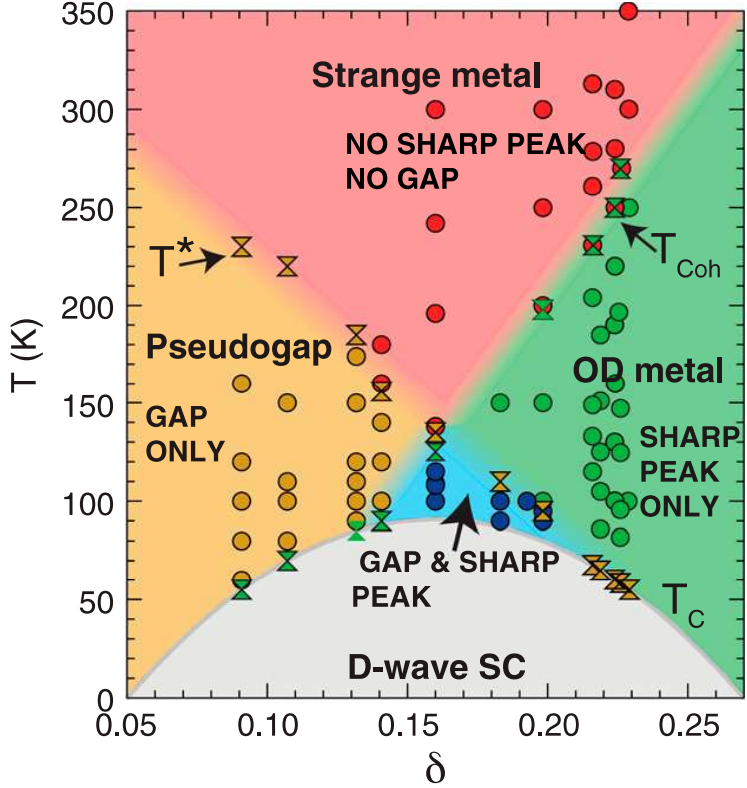}
\end{minipage}\hfill
\begin{minipage}[c]{0.28\linewidth}
\caption{{\footnotesize The cuprate phase diagram based on either the presence or absence of spectral gap and/or sharp peak. The latter indicates the existence of well-defined quasiparticle excitations, e.g., as in the more conventional metallic state in the overdoped region (From Ref.\onlinecite{UChatterjee}).}\label{fig.ARPESPhaseDiagram}}
\end{minipage}
\end{center}
\end{figure}

The spectral weight of the antinodal coherence peak, that marks the onset of superconductivity, decreases with decreasing doping \cite{JCCampuzano1}. This behaviour is unlike that in any conventional superconductors \cite{PALee2006}. A more recent and very detailed ARPES study \cite{UChatterjee} gives an overall perspective (Fig.\ref{fig.ARPESPhaseDiagram}) of the cuprate phase diagram based on the existence of gap and/or sharp spectral peak in the ARPES spectra. For example, the strange metal phase is characterized as the one with neither gap nor sharp peak, whereas the pseudogap has a gapped spectrum but no sharp peak. Interestingly, there is a narrow region above the superconducting dome around optimal doping having both gap and sharp spectral peak, similar to the superconducting state below $T_c$. 

{\bf 2. Tunneling spectroscopy:}  In scanning tunneling spectroscopy (STS) or scanning
tunneling microscopy (STM) one measures the tunneling current that flows between a sharp metallic tip and typically a 
conducting sample separated by thin insulating barrier, generally vacuum \cite{MTinkham1,OFischer1}. STM allows one to probe the local density of states (LDOS) $\rho(\omega=eV,\mathbf{r})$ at position $\mathbf{r}$ directly from
the differential tunneling conductance $dI/dV$ obtained by measuring tunneling current $I(V,\mathbf{r})$ due
to applied voltage difference $V$ between the tip and the sample
i.e.~$dI(V,\mathbf{r})/dV\propto\rho(\omega=eV,\mathbf{r})$.

\begin{figure}[hbt]
\begin{center}
\begin{minipage}[c]{0.58\linewidth}
\includegraphics[width=\linewidth]{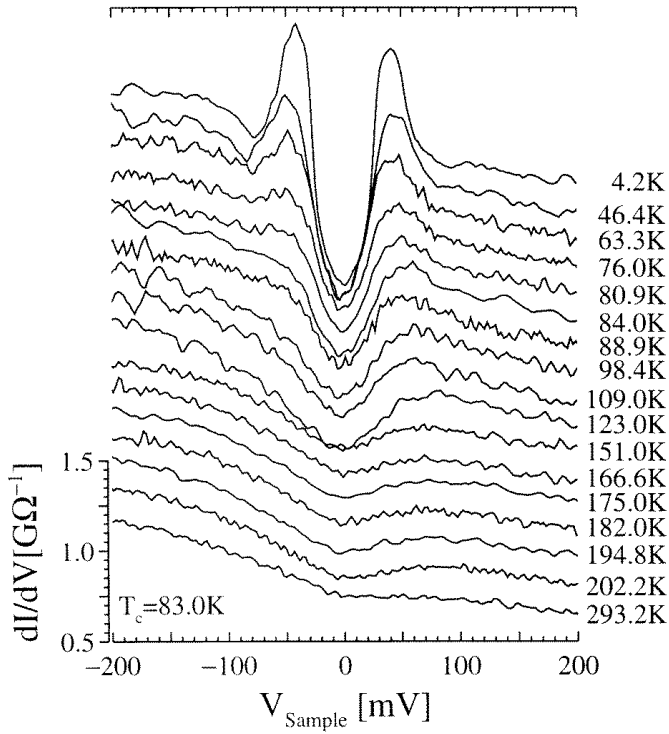}
\end{minipage}\hfill
\begin{minipage}[c]{0.38\linewidth}
\caption{{\footnotesize LDOS measured via differential tunneling conductance for an underdoped Bi2212 with $T_c=83$ K and pseudogap temperature $T^*\simeq 300 K$). A gap-like feature indicated by the suprresion of LDOS at low energies is seen even above $T_c$ till $T^*$ indicating a pseudogap. Two coherence peaks can be seen  below $T_c$. (From Ref.\onlinecite{CRenner1}).}\label{fig.STM_TDependence}}
\end{minipage}
\end{center}
\end{figure}

STM provides direct real-space information and hence is complementary to ARPES. By virtue of probing 
the local density of states with very high energy resolution, STM gives direct information \cite{OFischer1} about the gap in 
the excitation spectrum as well as other spectroscopic features such as scattering rate etc. The Pseudogap was first detected in tunneling spectroscopy by Tao {\it et al.} \cite{HJTao1}. A gap-like suppression of LDOS was found in the tunneling conductance of Bi2212 in the normal state. STM studies \cite{CRenner1} on underdoped Bi2212 have again shown a disappearance of the coherence peaks at $T_c$ and the existence of a V-shaped LDOS suppression in the normal state up to room temperature (see Fig.\ref{fig.STM_TDependence}). This feature 
can be identified with the signature of pseudogap. 

\begin{figure}[hbt!]
\begin{center}
\begin{minipage}[c]{0.58\linewidth}
\includegraphics[width=\linewidth]{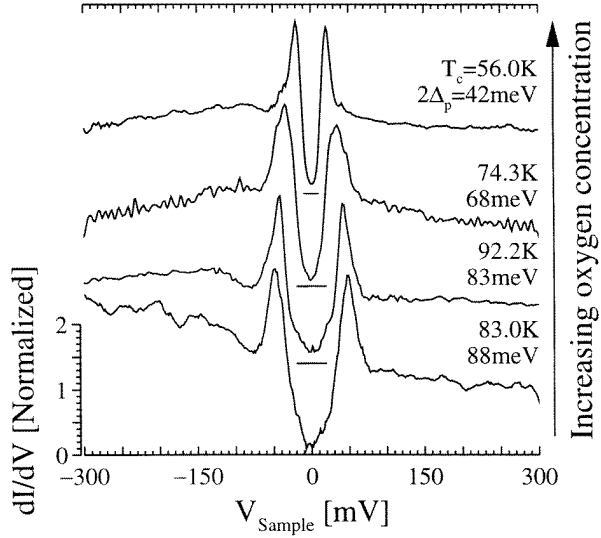}
\end{minipage}\hfill
\begin{minipage}[c]{0.38\linewidth}
\caption{{\footnotesize Doping dependence of LDOS measured through STS at low temperature. The separation between the peaks $2\Delta_p$ quantifies a superconducting gap that decreases with $x$. (From Ref.\onlinecite{CRenner1}).}\label{fig.STM_xDependence}}
\end{minipage}
\end{center}
\end{figure}

\begin{figure}[hbt!]
\begin{center}
\begin{minipage}[c]{0.58\linewidth}
\includegraphics[width=\linewidth]{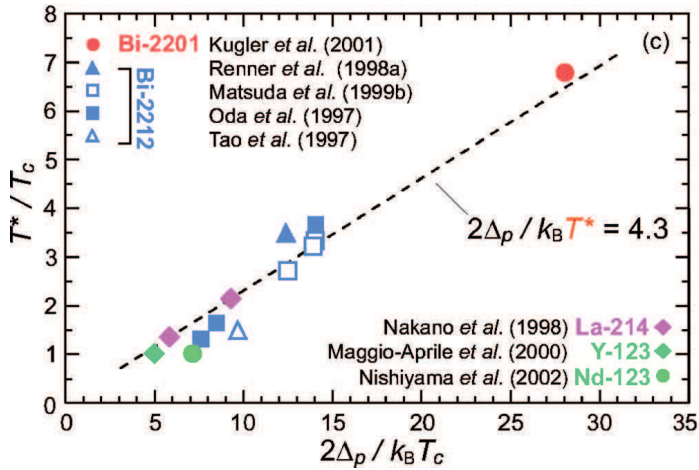}
\end{minipage}\hfill
\begin{minipage}[c]{0.38\linewidth}
\caption{{\footnotesize $T^*/T_c$ as a function of $2\Delta_p /k_BT_c$ obtained for various cuprates from STS. Dashed line corresponds to the mean-field $d$-wave relation $2\Delta_p /k_BT^*=4.3$, but with $T_c$ replaced 
by pseudogap temperature scale $T^*$. (From Ref.\onlinecite{MKugler1}, see Ref.\onlinecite{OFischer1} for references within the figure).} \label{fig.STM_GapTstar}}
\end{minipage}
\end{center}
\end{figure}

The magnitude of the gap in the superconducting state, measured via the separation between the peaks, is effectively temperature independent, unlike that of a BCS gap which closes as the temperature is raised through $T_c$. The same phenomenology applies to the magnitude of the pseudogap which, as in
ARPES, fills in with temperature but retains its magnitude up to room temperature in underdoped samples. The magnitude of the gap however does decrease with increasing $x$ (Fig.\ref{fig.STM_xDependence}). This is just the opposite to what is expected from BCS theory when the dome-shaped doping dependence of $T_c$ is taken into account (Fig.\ref{fig.EDC} {\bf f}).

STM reveals several important relations between the superconducting gap below $T_c$ and pseudogap above $T_c$ and between $T_c$ and $T^*$ \cite{MKugler1}.Firstly, the superconducting gap and pseudogap has the same scaling with doping, i.e.~increasing with decreasing doping. However the ratios  of these two gaps are nearly the same over different compounds and doping in spite of the fact that their absolute magnitudes change substantially between different compounds and doping. Also the ratio $2\Delta_p/k_\mathrm{B}T_c\approx 4.3$ amazingly follows the BCS $d$-wave relation \cite{HWon1} with $T^*$ replacing $T_c$ (Fig.\ref{fig.STM_GapTstar}). Here $\Delta_p$ denotes half of the separation between two superconducting peaks.

\section{Phenomenological theory of superconducting and pseudogap states} \label{sec.GLTheory}
A variety of approximate microscopic theories based on many different scenarios and ideas, from both strong and weak coupling perspectives, have 
been proposed to understand the cuprate phase diagram, and numerical techniques at various levels of sophistication have been
tried to address the issue of strong correlations in Hubbard and $t-J$ models. On the other hand, in
conventional superconductors and, in general, for the study of phase transitions, phenomenological
Ginzburg-Landau (GL) functionals \cite{PMChaikinBook} written down from very general symmetry grounds have provided useful 
descriptions for a variety of systems \cite{ISAranson2002}. Specially, Ginzburg-Landau theory has been proven to be 
complementary to BCS theory for attacking a plethora of situations in superconductors, e.g., inhomogeneities, structures of
isolated vortex and vortex lattice etc. \cite{MTinkham1}. In the same spirit, some of us have proposed and developed a phenomenological approach \cite{SBanerjee2011_1} for the 
superconducting and pseudogap states of the cuprates. 

In the phenomenological theory we attribute the pseudogap to short-range local spin-singlet pairing. This is in line with a large number of theories for cuprate superconductivity within the so-called ``preformed pair scenarios" \cite{MRNorman1}, especially those from the strong coupling point of view \cite{PALee2006}. We discuss below a possible microscopic underpinning of our theory from a strong correlation picture. However, we emphasize that we do not tie ourselves down to any particular microscopic calculation. Our theory essentially has two main empirical inputs, namely
\begin{enumerate}
\item A pairing temperature scale $T^*(x)$, mimicking the pseudogap temperature scale (Fig.\ref{fig.CupratePhaseDiagram}), below which local pairing amplitude becomes substantial.
\item A `bare' superfluid density $\rho_s\propto x$, that linearly increases with doping for small $x$, as implied by the Uemura correlation \cite{YJUemura1989} discussed before.
\end{enumerate}

As we discuss below, based on the above basic ingredients built into the proposed GL functional,  we can encapsulate a large number of well known cuprate phenomenologies in the form of a low-energy effective lattice functional of complex spin-singlet pair amplitudes that resides on the Cu-Cu bonds of the
$\mathrm{CuO_2}$ planes of cuprates. We summarize our results for average pairing amplitude in the superconducting and pseudogap phases, superfluid density, specific heat, vortex properties , fluctuation diamagnetism as well as thermoelectric transport quantity like Nernst coefficient \cite{SBanerjee2011_1,KSarkar2016,KSarkar2017} obtained using the proposed functional and compare them successfully with experiments. 

Motivated by the picture that emerges from our GL-like approach, we have also extended \cite{SBanerjee2011_2} it to obtain the 
electron spectral density that inevitably results from coupling between electrons and Cooper pair fluctuations. These results, and their implications for low energy electronic spectra measured through ARPES and STM, are also discussed. 

\subsection{Phenomenological GL-like functional}

\begin{figure}[hbt]
\begin{center}
\begin{minipage}[c]{0.58\linewidth}
\includegraphics[width=\linewidth]{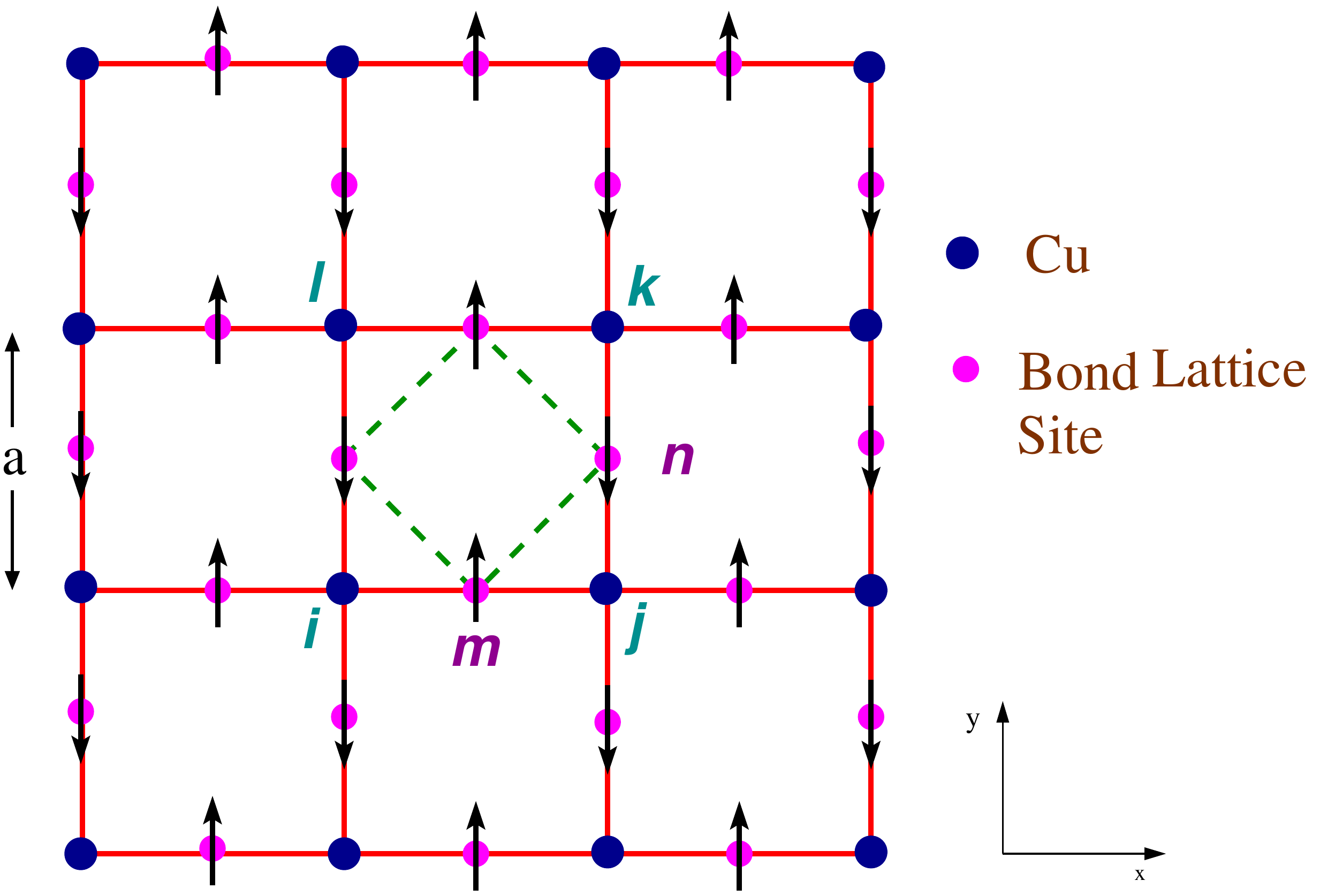}
\end{minipage}\hfill
\begin{minipage}[c]{0.38\linewidth}
\caption{{\footnotesize The square Cu lattice sites in the $\mathrm{CuO}_2$ plane and construction of the 
bond-centre lattice out of the centers of the Cu-O-Cu bonds. The arrows indicate the direction of equivalent
planar spins, with ${\bf S}_m=(\Delta_m \cos\phi_m,\Delta_m\sin \phi_m)$ representing the complex order
parameter $\psi_{ij}\equiv \psi_m=\Delta_m \exp(i\phi_m)$ and antiferromagnetic ordering of spins
translating into a long-range $d$-wave superconducting order.}\label{fig.BondLattice}}
\end{minipage}
\end{center}
\end{figure}

Based on our central assumption of the pairing origin of pseudogap, we propose that, as a minimal description, the free energy of a cuprate in the superconducting and pseudogap states can be expressed as a functional solely of the complex pair amplitude. We hypothesize a lattice free-energy functional similar in structure to the conventional GL functional \cite{MTinkham1}, but defined on the $\mathrm{CuO}_2$ square lattice. 
Fig.\ref{fig.BondLattice} shows the square planar lattice schematically. The free energy is assumed to be a functional of the complex spin-singlet pair amplitude
$\psi_{ij}\equiv\psi_m=\Delta_m\exp{(i\phi_m)}$, where $i$ and $j$ are nearest-neighbour sites of the square
planar Cu lattice and $m$ labels the bond-center lattice site located at the bond between the lattice site $i$ and $j$ (see Fig.\ref{fig.BondLattice}). The highly anisotropic cuprate superconductivity is modelled as a weakly coupled stack of $\mathrm{CuO_2}$ planes and we ignore, as a first approximation, the interplane coupling. The free-energy functional for
a single plane is assumed to have the lattice version of the conventional GL functional \cite{MTinkham1}, i.e.

\begin{eqnarray}
&&\mathcal{F}(\{\Delta_m,\phi_m\})=\sum_m \left(A\Delta_m^2 + \frac{B}{2}\Delta_m^4\right)+C \sum_{<mn>}  \Delta_m \Delta_n \cos(\phi_m-\phi_n),
\label{Eq.functional}
\end{eqnarray}
 We posit the above lattice functional to be more fundamental, in principle, than the conventional continuum GL functional due to underlying strongly correlated nature of the cuprate superconductors and owing to the fact that they have very short coherence length ($\sim 15-20 \AA$).  A Gor'kov like \cite{Gorkov1959} microscopic interpretation of  $\psi_{ij}$ is as an average spin-singlet nearest-neighbor Cooper pair
amplitude, i.e., $\psi_{ij}\equiv\langle \hat{b}_{ij} \rangle/\sqrt{2}\equiv(1/2)\langle a_{i\downarrow}a_{j\uparrow}-a_{i\uparrow}a_{j\downarrow}\rangle$. The sites $i$
and $j$ are different because strong electron repulsion $U$ disfavors on-site  pairing, while the existence of large nearest-neighbor antiferromagnetic spin-spin interaction in the 
parent (undoped) cuprate is identically equivalent for 
spin-$\frac{1}{2}$ electrons to attraction between nearest-neighbor pairs, i.e. $J (\hat{\mathbf{S}}_i.\hat{\mathbf{S}}_j-\frac{1}{4}\hat{n}_i\hat{n}_j) = -J \hat{b}^\dagger_{ij} \hat{b}_{ij}$. This favours the formation of nearest-neighbour bond spin-singlet pairs. 

The `on-site' part of the above functional [eq.\eqref{Eq.functional}] has been written down in a bare minimum form, as a sum of quadratic and quartic terms in $\Delta_m$ in the image of conventional GL theory, for simplicity. We remark that unlike in the original GL theory where the parameters $A$ and $B$ are proposed near $T_c$, we propose this simple form for all $T$ and all $x$, as a qualitatively correct one. The parameters $A$ and $B$ depend, in general, on $x$ and $T$. We assume that $B$ is a positive constant independent
of $x$ and $T$ and choose $A(x,T)\propto (T-T^*(x))$ to change sign along a straight 
line $T^*(x)=T_0(1-x/x_c)$ running from $T= T_0$ at $x=0$ to $T=0$ at $x=x_c$. As a first approximation, this line can be 
identified with the pseudogap temperature $T^*(x)$ because the magnitude of the local pair ampliitude, $\langle\Delta_m\rangle$, can increase 
dramatically as the temperature crosses
this line due to sign change of $A$. 

The occurrence of superconductivity in the model is characterized by a nonzero stiffness for long-wavelength phase fluctuations and is determined by the phase coupling term $C$ [eq.\eqref{Eq.functional}]. The parameter $C$ is taken to be proportional to $x$ in conformity with the Uemura correlation \cite{YJUemura1989}. Also, $C>0$ so that the zero-temperature superconducting state has $d$-wave symmetry [see Fig.\ref{fig.BondLattice}]. We discuss the choice of doping and temperature dependences of the parameters $A$, $B$ and $C$ in more detail in Appendix \ref{app.GLParameters}. 

{\bf Effect of magnetic field:} An applied magnetic field perpendicular to the $\mathrm{CuO_2}$ planes giving rise to, e.g., the orbital magnetization and magnetothermoelectric effects \cite{KSarkar2016,KSarkar2017}, is incorporated in our model through a bond flux $A_{mn}=(2\pi/\Phi_0)\int_m^n\mathbf{A}\cdot d\mathbf{r}$, which modifies $(\phi_m-\phi_n)$ to $(\phi_m-\phi_n-A_{mn})$ in eq.\eqref{Eq.functional}, as usual in a lattice GL functional. Here $\mathbf{A}$ is the vector potential. Assuming extreme type-II limit as appropriate for the cuprate superconductors, the magnetic field $H$ is given by the condition $\sum_{\Box} A_{mn} = \Phi$, where $\sum_{\Box}$ is a sum over a plaquette of the lattice and $\Phi=Ha^2/\Phi_0$ is the magnetic flux per plaquette in units of the universal flux quantum.

{\bf Effect of quantum phase fluctuations:} As we discuss below, the superconducting transition temperature $T_c$ obtained in our model follows a dome-shaped curve as a function of $x$. However, on the extreme underdoped and overdoped regimes the superfluid density becomes small in our model \cite{SBanerjee2011_1}. In these regimes, one needs to take into account the effect of quantum phase fluctuations. These would renormalize $T_c$ to zero at finite doping in the underdoped side and their importance is well-supported by experiments~\cite{Broun2007,Hetel2007} and theoretical analysis~\cite{Franz2006}. We can incorporate quantum phase fluctuation effects in our formalism \cite{SBanerjee2011_1} by supplementing the free-energy functional of eq.\eqref{Eq.functional} with the following term 
\begin{eqnarray}
\mathcal{F}_Q(\{\hat{q}_m\})&=&\frac{1}{2}\sum_{mn} \hat{q}_m V_{mn} \hat{q}_n \label{Eq.functionalQ}
\end{eqnarray}
Here $\hat{q}_m$ is the Cooper pair number operator at site $m$, and $\phi_m$ in eq.\eqref{Eq.functional} should be treated as a quantum mechanical operator $\hat{\phi}_m$, canonically conjugate to $\hat{q}_m$ so
that $[\hat{q}_m,\hat{\phi}_n]=i\delta_{mn}$. We take the simplest possible form for $V_{mn}$ i.e.~$V_{mn}=V_0\delta_{mn}$, where $V_0$ is the strength of on-site Cooper pair interaction, an additional phenomenological parameter that can fixed such that  $T_c(x)$ curve computed in the phenomenological theory matches the experimental one \cite{SBanerjee2011_1}. We do not include these quantum fluctuations
explicitly for calculating thermodynamic quantities, e.g.~specific heat, magnetization etc., associated with pairing fluctuations, since quantum phase fluctuations bring about qualitative changes only at the extreme end of the dome on the underdoped side. For other values of $x$, these fluctuations only renormalize the values of the parameters $A$, $B$ and $C$ of our functional \eqref{Eq.functional}. We assume that such renormalizations have already been taken into account while choosing these parameters in tune with experiments. Next we discuss some general aspects of our model and its possible microscopic origin.
 
 {\bf General aspects of the phenomenological theory:}  As clearly evident from eq.\eqref{Eq.functional}, since $\Delta_m\Delta_n\cos(\phi_m-\phi_n)=-(|\psi_m-\psi_n|^2-\Delta_m^2-\Delta_n^2)$, the intersite term can be readily identified with the discretized version of usual spatial derivative term $|\nabla \psi|^2$. The model is thus of the form of a conventional GL model, albeit one that 
is defined on a lattice at the outset. As discussed above, defining the complex singlet pair amplitude to reside on the $\mathrm{Cu-Cu}$ bonds (Fig.\ref{fig.BondLattice}) and choosing the phase coupling term $C$ to be positive, we immediately get a $d$-wave superconducting order at low temperature. The negative sign of $C$ corresponds to more conventional $s$-wave BCS like case. For quantitative comparison with experiments, the lattice should be thought of more appropriately as a phenomenological one that emerges upon coarse graining, albeit with a coarse graining length only three to five times larger than microscopic lattice spacing $a$ \cite{KSarkar2016}. Hence, our model, in practice, can be thought of as the discretized version of a continuum theory with the lattice spacing $a$ as a suitable ultraviolet cutoff to describe long wavelength physics. In the presence of magnetic field $H$, the lattice constant is also equivalent to a field scale $H_0=\Phi_0/(2\pi a^2)$, defined through the flux quantum $\Phi_0=hc/2e$. In principle, the field scale $H_0$ can be obtained by fitting the field dependence of magnetization with that of experiment \cite{KSarkar2016}.

 The free-energy functional in eq.~\eqref{Eq.functional} can also be viewed as the Hamiltonian of an XY model with fluctuations in the magnitude of `planar spin' $\psi_m$, where the on-site term in eq.\eqref{Eq.functional} simply controls the temperature and doping dependence of the magnitude. The form of the free-energy functional might seem superficially similar to the widely used model of granular superconductors \cite{Ebner1981}. However, we would like to re-emphasize that we do not assume any underlying granularity of our system, as mentioned above.  Such phenomenological lattice models, in the extreme $XY$ limit, have been employed in the past to study superconductivity in non-granular lattice systems, especially in the context of cuprates~\cite{Carlson1999,Paramekanti2000,Franz2006,Podolsky2007}.

Additionally, even though the form of our functional is mainly
motivated by cuprate phenomenology, as we discuss below, a similar functional arises quite naturally in a strong correlation framework for a doped Mott insulator \cite{Baskaran1988,Drzazga1989,Drzazga1990}. In general, in such a functional, the single-site term of eq.\eqref{Eq.functional} will have more complicated form\cite{Drzazga1989,Drzazga1990}, having many terms in a power series expansion of $\Delta_m$, in addition to the quadratic and quartic ones that our functional does. But, as we discuss below, the superconducting dome is reproduced quite reasonably by truncating the functional to quartic order. In addition, several other experimentally observed thermodynamic properties of the cuprates over the entire pseudogap regime are also reproduced by this simplified form of the functional~\cite{SBanerjee2011_1}.

{\bf Microscopic underpinning:} GL theories for cuprates have been proposed by a large number of authors, arising either out of a particular model for electronic behavior and often coupled with the assumption of a particular
`glue' for binding electrons into pairs \cite{Baskaran1988,Drzazga1989,LTewordt1989}, or out of lattice symmetry 
considerations \cite{DLFeder1997,AJBerlinsky1995}. The functional in eq.\eqref{Eq.functional} is consistent with square lattice 
symmetry and, in principle, does not assume any particular electronic approach, e.g. weak coupling 
or strong correlation. However, some of the properties of the coefficients are natural in a strong electron correlation framework. For example, mobile holes in such a system can cause a transition between a state in which there is a Cooper pair in the $x$ directed $ij$ bond (Fig.\ref{fig.BondLattice}) to one in which the Cooper pair is in an otherwise identical but $y$ directed 
bond $jk$ nearest to it (or vice versa), thus leading to a nonzero intersite phase coupling term  in
eq.\eqref{Eq.functional}. This is probably connected with the observed~\cite{EPavarini2001} empirical correlation 
between $T_c$ and the diagonal or next-nearest-neighbor hopping amplitude $t'$ of electrons in the Cu lattice. 

A phenomenological approach such as ours does not point to a specific `glue' which binds electrons on nearest-neighbor 
sites into spin singlet pairs. However, as mentioned before, in a strong 
correlation picture, a natural source is $J$. In general therefore, the nearest-neighbor interaction can be thought of as a mixture of antiferromagnetic spin exchange and pair attraction. As the Mott insulator is doped and the holes become mobile, it seems likely (from experiment) that the interaction is more conveniently described as latter
than the former in a mean-field sense. A number of microscopic variational $T=0$ calculations  appropriate
for competing spin density wave and $d$-wave superconducting ground states as a function of hole doping exist
in the literature~\cite{MOgata2008,TGiamarchi1991,AHimeda1999}; they generally suggest a magnetic ground state or a 
coexistence of magnetism and superconductivity for low hole doping and a superconducting ground state for higher doping.  
Here, we consider only the pair degrees of freedom and disregard the spin. So, if this description is appropriate, 
the limiting ($x\rightarrow 0$) pseudogap or bound pair energy scale $T_0$ is expected to be of order $J$ from which the pair
attraction originates in the strong correlation picture. Nevertheless, $T_0$ is a factor of two or three smaller than $J$. The
decrease of the pseudogap energy scale with $x$ is presumably due to holes because their presence implies at the very least a decrease in the number of nearest-neighbor electron pairs that can be formed. Some early results \cite{PWAnderson2002}
suggest a mean-field `effective' $J_\mathrm{eff}\simeq J-xt$ where $t$ is the nearest-neighbor hopping. 
This may be relevant to the linear decrease of $T^*(x)$ with doping $x$.

\subsection{Thermodynamic quantities obtained from the phenomenological theory}

The main objective of our phenomenological approach is to investigate whether the free energy functional defined in eq.\eqref{Eq.functional} provides a good description of experimental results over a wide range of $x$ and $T$. To this end, we have calculated large number of thermodynamic quantities associated with pairing fluctuations \cite{SBanerjee2011_1,KSarkar2016} and directly compared them with the experimental observations discussed in section \ref{sec.cupratephenomena}. Here we list some of the quantities computed within the phenomenological theory. The details of the calculations and the results can be found in Refs.\onlinecite{SBanerjee2011_1,KSarkar2016,KSarkar2017}. In section \ref{Sec.Results}, we summarize our main results and briefly discuss how these compare with the cuprate phenomena discussed in section \ref{sec.cupratephenomena}.

We have calculated the zero-temperature gap $\Delta_0(x)$ and superfluid density $\rho_s^0(x)$ as a function of doping by minimizing the free-energy functional \eqref{Eq.functionalQ} with respect to pairing amplitude $\Delta_m$ at $T=0$ for the uniform $d$-wave state shown in Fig.\ref{fig.BondLattice}. Thermodynamic quantities, such as average local pairing amplitude $\langle \Delta_m\rangle$, specific heat $C_v$ and orbital magnetization $M$ etc., are calculated via canonical thermal averages, $\langle \dots\rangle=(1/Z)\int \prod_m d\phi_m d\Delta_m (\dots) e^{-\beta \mathcal{F}(\Delta_m,\phi_m)}$, where $Z$ is the partition function. The calculations have been done via standard techniques like mean-field theory, Monte-Carlo (MC) simulation etc. We have also calculated properties, e.g. core-structure and core-energy $E_c(x)$, of topological defects, i.e. a superconducting vortex, via numerical optimization with appropriate boundary condition \cite{SBanerjee2011_1}. The vortex core-energy has important implications for the transport coefficients in the underdoped regime, as discussed in Ref.\onlinecite{KSarkar2016}.  

\section{Extensions of the phenomenological theory to describe thermoelectric transport and low-energy electronic spectra} \label{sec.extensions}
We have extended the phenomenological theory in several ways, e.g., to study the contributions of pairing fluctuations to thermoelectric transport, such as Nernst coefficient, and also to look into the effects of these fluctuations on low-energy electronic spectral density as directly probed  via ARPES and STM.  We discuss these extensions below.

\subsection{Relaxation dynamics using the phenomenological theory}
To study the influence of pairing fluctuations on transport properties we implement\cite{KSarkar2017} ``model A'' dynamics \cite{PMChaikinBook} for a time-dependent complex pair amplitude $\psi_m(t)=\Delta_m(t)e^{i\phi_m(t)}$ given by the stochastic equation
\begin{equation}\label{Eq.TDGL}
\tau D_t\psi_m(,t)=-\frac{\partial \mathcal{F}(\lbrace \psi_m, \psi_m^{*}\rbrace)}{\partial \psi_m^{*}(t)}+\eta_m(t).
\end{equation}
Here $\mathcal{F}(\lbrace \psi, \psi^{*}\rbrace)$ is a the free energy functional [eq.\eqref{Eq.functional}], that incorporates the effect of electromagnetic field in the inter-site phase coupling term, as discussed earlier, via a vector potential $\mathbf{A}_{mn}(t)$, which can be time-dependent in general. We define a covariant time derivative $D_t=(\frac{\partial}{\partial t}+iV_m(t))$, where $V_m(t)$ is a scalar potential at site $m$. The time scale $\tau$, which provides the characteristic temporal relaxation scale of the order parameter dynamics, can in general be complex. However it is required to be real in the presence of particle-hole symmetry, that translates to the requirement that the equation of motion for $\psi_m^{*}$ be the same as for $\psi_m$ under the simultaneous transformation of complex conjugation ($\psi_m\rightarrow \psi_m^{*}$) and magnetic field inversion ($H\rightarrow -H$). Evidence of particle-hole symmetry in the form of no appreciable Hall or Seebeck effect is seen in the experimentally accessible regime of the superconductors we study here, and thus we take $\tau$ to be real in our calculations. The thermal fluctuations are introduced through  $\eta_m(t)$  with the complex Gaussian white noise correlator $\langle \eta^*_n(t)\eta_m(t')\rangle=2k_\mathrm{B}T\tau \delta_{mn}\delta(t-t')$.

The dynamical model Eq.~\ref{Eq.TDGL} is the simplest one which yields an equilibrium state in the absence of driving potentials. For conventional superconductors, it can be derived microscopically within BCS theory above and close to the transition temperature $T_c$. However, it has been used phenomenologically to study transport previously in situations where the microscopic theory is not known, such as for the cuprates~\cite{Ussishkin2002,Mukerjee2004,Podolsky2007}. We employ the model in a similar spirit here.

{\bf Transport coefficients:} To compute transport coefficients, we numerically integrate the dynamical equations \eqref{Eq.TDGL}, in the presence of either temperature gradient $\nabla T$ or electric field $\mathbf{E}$, applied via suitable choice of gauge potentials $A_{mn}(t)$ and $V_m(t)$. The transport current densities, obtained in our model after appropriate subtractions of magnetization currents \cite{KSarkar2017}, can be related to $\nabla T$ and $\bf{E}$ using standard linear response relations
\begin{center}
$\begin{pmatrix} {\bf J}_{\mathrm{tr}}^e \\ {\bf J}_{\mathrm{tr}}^Q \end{pmatrix} = \begin{pmatrix} \hat{\sigma} & \hat{\alpha} \\ \ \hat{\tilde{\alpha}} & \hat{\kappa} \end{pmatrix}   \begin{pmatrix} \bf{E} \\ -\nabla{T} \end{pmatrix}$,
\end{center}
where $\hat{\sigma}$, $\hat{\alpha}$, $\hat{\tilde{\alpha}}$, $\hat{\kappa}$ are the electrical, thermoelectric, electro-thermal and thermal conductivity tensors, respectively. We have mainly studied the Nernst signal $e_N$, which is the electric field ($E_y$) response to a temperature gradient ($\nabla_x T$) in the direction transverse to the electric field in the presence of a
magnetic field $H$ along $z$-direction, i.e.~$e_N\equiv E_y/\nabla_x T$ with open circuit boundary conditions. Under reasonable assumptions \cite{KSarkar2017}, the Nernst coefficient, $\nu=e_N/H$ can be connected to the thermoelectric coefficients via the simple relation $\nu=\alpha_{xy}/(H\sigma_{xx})$, that we use. We summarize our results of Nernst coefficient calculated over the whole doping-temperature range in section \ref{Sec.Results}. 

\subsection{Pairing fluctuations and low-energy electronic spectra in cuprates} \label{sec.spectral}

Decipherment of the physics of high-T$_c$ cuprates should necessarily involve the understanding of
electronic excitations and properties related to them in the various parts of cuprate phase diagram
(Fig.\ref{fig.CupratePhaseDiagram}). Hence a phenomenological theory, such as ours, which aims to describe the pseudogap phase as one consisting of preformed pairs of \emph{a specific type}, is required to include both electrons and Cooper pairs 
of the same electrons coexisting and \emph{necessarily} coupled with each other. In our GL-like approach only the latter are 
explicit, while the former are \emph{integrated out}. However, effects connected with the pair degrees of freedom are often explored via their coupling to electrons, e.g. via ARPES and STM, as discussed earlier. Here we develop a unified 
theory of electronic excitations in the superconducting and pseudogap phases, using a model of electrons coupled to spatially and temporally fluctuating nearest-neighbor bond singlet Cooper pairs. We discuss the theory, its basic principles and assumptions here, and a number of its
consequences with respect to cuprate phenomena are summarized in section \ref{Sec.Results}.

{\bf Basic notions and assumptions behind the approach:}
The main difficulty in a theory of the above kind is the description of electrons in a presumably strongly
correlated system such as a cuprate, which is perhaps best described as a doped Mott insulator \cite{PALee2006} with strong low-energy antiferromagnetic correlation between electrons as well as various other types long-and short-range ordering tendencies. This is the major problem for constructing a microscopic theory in the entire field of strongly correlated electrons and dealing with such problems in a controlled manner would probably continue to be an active field of research and one of the most challenging tasks in condensed matter physics for years to come .

However, irrespective of the microscopic origin, the interacting pair degrees of
freedom probably still can be described by a phenomenological theory of the kind described here. To pursue this path, one needs to commit to some kind of model for 
electron dynamics which therefore implies an approach to the coupling between single electronic and pair
degrees of freedom. We develop what we believe is a minimal theory, appropriate for low-energy physics. We assume that for 
energies $|\omega|\aplt \Delta$ ($\Delta$ is of the order of zero temperature superconducting gap), well defined 
electronic states, namely tight binding lattice states with a renormalized hopping/bandwidth exist and couple to low energy 
pair fluctuations. Superconducting ordering/phase stiffness and fluctuations are reflected in the pair-pair correlation 
function. This correlation function has a generic form, especially near $T_c$, which, for example, may be obtained from the 
GL-like functional [eq.\eqref{Eq.functional}] or, as we show in Ref.\onlinecite{SBanerjee2011_2}, can be deduced from various experiments. The \emph{inevitable} coupling between electrons and low lying fluctuations of pairs, made of
the same electrons leads to a self energy with a 
significant structure for electron momentum $\mathbf{k}$ along the Fermi surface and low electronic excitation energy 
$\omega$. 

Physically, we have electrons (e.g.~those with energy near the Fermi energy) moving in a medium of evanescent 
pairs which have finite range $d$-wave correlation for $T>T_c$ and have long range order of this kind for $T<T_c$ in 
addition to `spin wave' like fluctuations. On approaching $T_c$ from above, collective $d$-wave symmetry superconducting
correlations develop among the pairs with a characteristic superconducting coherence length scale $\xi$ which
diverges at the second-order transition temperature $T_c$. These correlations have a generic form at large distances
($>>$ the lattice spacing $a$). Our theory captures the effects of these correlations on low-energy unpaired electronic degrees of freedom in the superconducting and pseudogap phases.

{\bf Model:}  To motivate the microscopic coupling between the spin-singlet bond pair and electrons, we take the $tJ$ model \eqref{eq.tJModel} as the basic starting point. However to circumvent the daunting task of treating the no double occupancy constraint discussed in section \ref{sec.materials}, we use the following simplified Hamiltonian, the so-called `plain vanilla' \cite{PWAnderson2004} version of $tJ$ model, i.e.
\begin{eqnarray}
\hat{\mathcal{H}}&=&-\sum_{ij,\sigma} \tilde{t}_{ij} a^\dagger_{i\sigma} a_{j\sigma} - \tilde{J}\sum_{<ij>} \hat{b}^\dagger_{ij}\hat{b}_{ij} \label{eq.Hamiltonian_Pair}.
\end{eqnarray}
Here the effective hopping $\tilde{t}_{ij}=g_t t_{ij}$ and effective antiferromagnetic exchange $\tilde{J}=g_sJ$, which acts as a pair attraction, are strongly affected by correlations via single-site Gutzwiller renormalization factors $g_t=2x/(1+x)$ and $g_s=4/(1+x^2)$ \cite{PWAnderson2004}. The bare hopping amplitude $t_{ij}$ involves the nearest neighbor ($t$), next-nearest neighbor 
($t'$) and further neighbor ($t''$) hopping terms. In our calculations, we use the above mentioned homogeneous 
Gutzwiller approximation with standard values for $t_{ij}$ ($t=300$ meV, $t'/t=-1/4$ and $t''=0$, see 
e.g.~\cite{AParamekanti1-6}). 
       
Very generally, assuming pairing fluctuations to be relevant at low energies, the bond-pair attraction term term eq.\eqref{eq.Hamiltonian_Pair}, can be written via Hubbard-Stratonovich transformation as a time and space dependent bond pair potential acting on electrons and characterized 
by a field $\psi_m(\tau)$ [$\tau$ is the imaginary time; $0<\tau<\beta=1/(k_BT)$]. The saddle point of the resulting action in the static limit gives rise to the conventional mean field approximation in which the second term in
eq.\eqref{eq.Hamiltonian_Pair} is written as 
\begin{eqnarray}
-\tilde{J}_{ij}(\langle \hat{b}^\dagger_{ij} \rangle \hat{b}_{ij}+\hat{b}^\dagger_{ij}\langle
\hat{b}_{ij}\rangle -\langle \hat{b}^\dagger_{ij}\rangle \langle \hat{b}_{ij}\rangle)
\label{eq.MFTDecoupling_ARPES}
\end{eqnarray}  
and the average $\psi_m=\psi_{ij}\equiv \tilde{J}_{ij}\langle \hat{b}_{ij} \rangle$ is determined self-consistently
(mean field theory).

The effective Hamiltonian we use is of the form of
eq.\eqref{eq.Hamiltonian_Pair} with the second term in it replaced by eq.\eqref{eq.MFTDecoupling_ARPES}. This describes 
two coupled fluids, namely a fermionic fluid and a bosonic fluid, represented respectively by the on-site
electron field $a^\dagger_{i\sigma}$ and the bond Cooper pair field $\psi_{ij}=\psi_m$. The properties of $\psi_m$ needed 
in our calculation are its mean value $\langle \psi_m \rangle$ (nonzero below $T_c$) and the fluctuation part of the 
correlation function $\langle \psi_m\psi^*_n \rangle$, whose {\it universal} form for large $|\mathbf{R}_m-\mathbf{R}_n|$ 
near $T_c$ is what we use. These arise from an inter-site term of the  form shown in eq.\eqref{Eq.functional}.

\begin{figure} [hbt]
\begin{center}
\begin{minipage}[c]{0.58\linewidth}
\includegraphics[width=\linewidth]{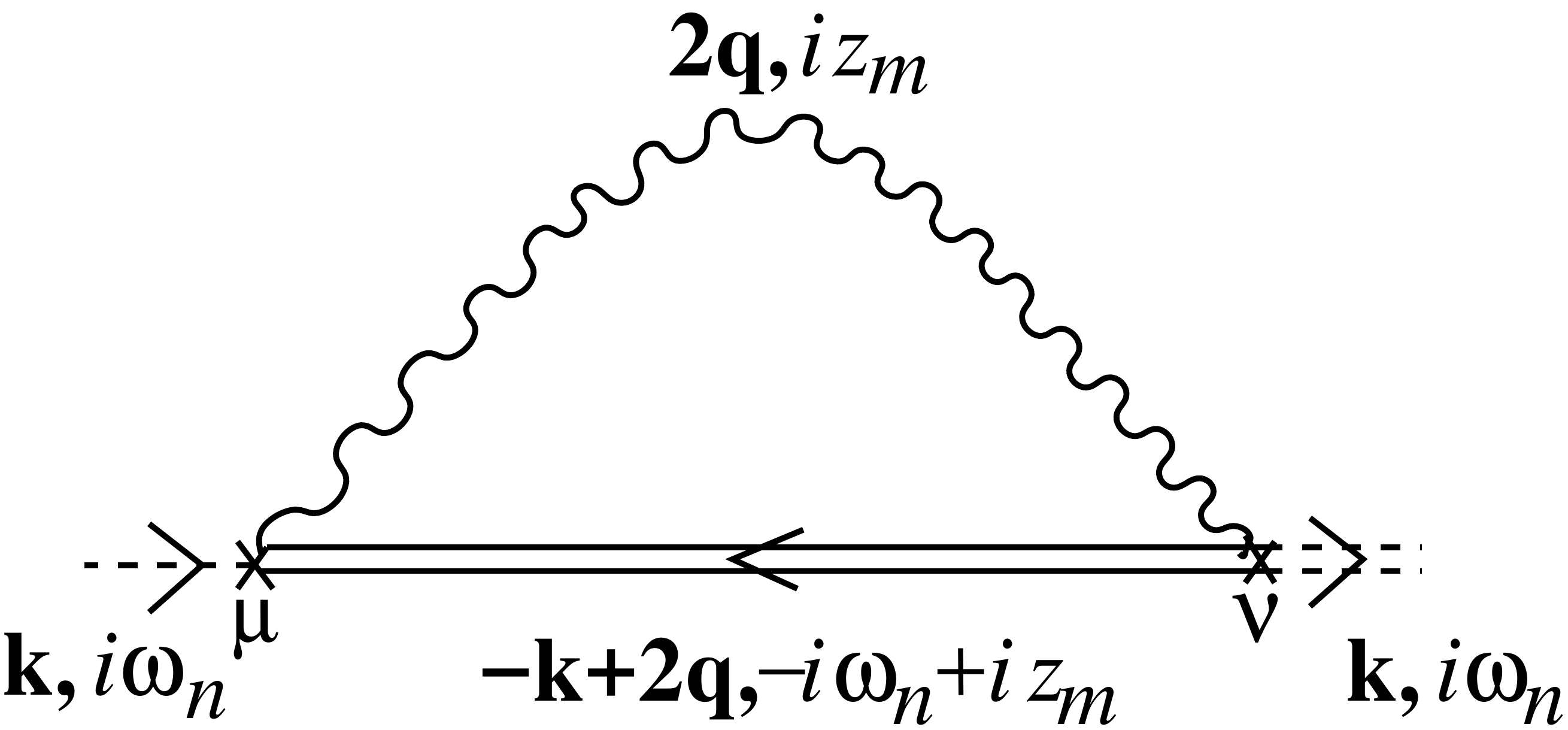}
\end{minipage}\hfill
\begin{minipage}[c]{0.38\linewidth}
\caption{{\footnotesize {\bf The pair fluctuation exchange process for the self energy of electrons.} 
Feynman diagram of the process in which an electron ($\mathbf{k}$,$i\omega_n$) virtually becomes 
another electron $(-\mathbf{k}+2\mathbf{q},-i\omega_n+iz_m)$ and absorbs a Cooper pair $(2\mathbf{q},iz_m)$ in the 
intermediate state. The curly line denotes the Cooper pair propagator $D_{\mu\nu}(2\mathbf{q},iz_m)$ 
(here $z_m=2m\pi/\beta$ is the bosonic Matsubara frequency, $m$ being an integer).}\label{fig.SelfEnergyApproximation}}
\end{minipage}
\end{center}
\end{figure}

For a translationally invariant system described by eq.\eqref{eq.Hamiltonian_Pair}, the electron Green's function
satisfies the Dyson equation
\begin{eqnarray}
G^{-1}(\mathbf{k},i\omega_n)&=&(G_0)^{-1}(\mathbf{k},i\omega_n)-\Sigma(\mathbf{k},i\omega_n),
\label{eq.DysonEquation}
\end{eqnarray}
where $\Sigma(\mathbf{k},i\omega_n)$ [$\omega_n=(2n+1)\pi/\beta$ is the fermionic Matsubara frequency with
$n$ as an integer] is the irreducible self energy, originating from the coupling between bond 
pairs and electrons with bare propagator $G_0(\mathbf{k},i\omega_n)=(i\omega_n-\xi_\mathbf{k})^{-1}$, 
$\xi_\mathbf{k}=\epsilon_\mathbf{k}-\mu$ and $\epsilon_\mathbf{k}$ is the Fourier transform of the hopping 
$\tilde{t}_{ij}=\tilde{t}_{\mathbf{i}-\mathbf{j}}$ ($\mu$ is the chemical potential). We use the well-known lowest-order bosonic fluctuation exchange
approximation for $\Sigma(\mathbf{k},i\omega_n)$ shown diagrammatically in Fig.\ref{fig.SelfEnergyApproximation}.

Close to $T_c$, the temporal decay of long-wavelength fluctuations is specially slow due to critical
slowing down \cite{PMChaikinBook}, so that they can be regarded as quasistatic. The self energy $\Sigma$ can then be expressed as \cite{SBanerjee2011_1}
\begin{eqnarray}
\Sigma(\mathbf{k},i\omega_n)&=&-\frac{1}{N}\sum_{\mathbf{q},\mu,\nu} G_0(-{\mathbf{k}+2\mathbf{q}},-i\omega_n)
D_{\mu\nu}(2\mathbf{q})f_\mu(\mathbf{k},\mathbf{q})f_\nu(\mathbf{k},\mathbf{q})\label{eq.StaticSelfEnergy}
\end{eqnarray}
where $N$ is the total number of Cu sites on a single $\mathrm{CuO_2}$ plane and $\mu$, $\nu$ refer to the direction of 
the bond i.e. $x$ or $y$ (see Fig.\ref{fig.BondLattice}). $D_{\mu\nu}(2\mathbf{q})=\sum_\mathbf{R} D_{\mu\nu}(\mathbf{R})
\exp{(-i2\mathbf{q}.\mathbf{R})}$ is the static pair propagator of Cooper pair fluctuations with
\begin{eqnarray} 
D_{\mu\nu}(\mathbf{R}_{i\mu}-\mathbf{R}_{j\nu})=\langle \psi_{i\mu}\psi^*_{j\nu}\rangle.
\label{eq.StaticPairCorrelator}
\end{eqnarray} 
The quantity $f_\mu(\mathbf{k},\mathbf{q})=\cos[(k_\mu-q_\mu)a]$ is a form factor arising from the coupling between 
a tight-binding lattice electron and a nearest-neighbour bond pair. Because of the $d$-wave long-range order (LRO) described as
`$\mathrm{Ne'el}$' order (Fig.\ref{fig.BondLattice}) of the `planar spin' $\psi_m$ in the bipartite bond-centre lattice, the 
standard sublattice transformation [i.e.~$\psi_m\rightarrow \tilde{\psi}_m=\Delta_m\exp{(i\tilde{\phi}_m)}$ where
$\tilde{\phi}_m=\phi_m$ for $x$-bonds and $\tilde{\phi}_m=\phi_m+\pi$ for $y$-bonds] implies
\begin{eqnarray}
D_{xx}(\mathbf{R})=D_{yy}(\mathbf{R})=-D_{xy}(\mathbf{R})=-D_{yx}(\mathbf{R})\equiv D(\mathbf{R}),
\end{eqnarray}
 where $D(\mathbf{R})$ can be written as
\begin{eqnarray}
D(\mathbf{R}_m-\mathbf{R}_n)=\langle \tilde{\psi}_m\rangle \langle
\tilde{\psi}^*_n\rangle+\widetilde{D}(\mathbf{R}_m-\mathbf{R}_n)\,.
\label{eq.UrsellFunction}
\end{eqnarray}
Here $\widetilde{D}(\mathbf{R})$ is the fluctuation term. The LRO part $\langle \tilde{\psi}_m\rangle\equiv \Delta_d$ leads 
to a $d$-wave Gor'kov like gap with $\Delta_\mathbf{k}= (\Delta_d/2)(\cos{k_xa}-\cos{k_ya})$; the corresponding electron 
self energy is $\Sigma(\mathbf{k},i\omega_n)=\Delta_\mathbf{k}^2/(i\omega_n+\xi_\mathbf{k})$. In widely used 
phenomenological analyses \cite{MRNorman1998} of ARPES data, this form is used above $T_c$ with lifetime effects, 
both diagonal and off-diagonal in particle number space added to $\Sigma$, i.e.~
\begin{eqnarray}
\Sigma(\mathbf{k},\omega)&=&-i\Gamma_1+\frac{\Delta_\mathbf{k}^2}{\omega+\xi_\mathbf{k}+i\Gamma_0}
\end{eqnarray} 
Here $\Gamma_1$ is single-particle scattering rate and $\Gamma_0$ is assumed to originate due
to finite life-time of \emph{preformed} $d$-wave pairs \cite{MRNorman1998}.

We propose here that as described above, the electrons move (above $T_c$) \emph{not} in a pair field with $d$-wave LRO which
decays in time at a rate put in by hand, but in a nearly static pair field with growing
correlation length $\xi$.  We assume, as appears quite natural for a system with characteristic length scale
$\xi$ that $\widetilde{D}(\mathbf{R})\sim \exp(-R/\xi)$ for large $R$, while $\xi$ diverges at $T_c$. 
This natural form for $\widetilde{D}(R)$ is found, for example, in the Berezinskii-Kosterlitz-Thouless (BKT) theory 
\cite{PMChaikinBook} for two dimensions and in a GL theory for 
all dimensions. For large correlation lengths, 
$\Sigma(\mathbf{k},i\omega_n)$ obtained in our theory is nearly the same as that for preformed $d$-wave symmetry pairs. 
Below $T_c$, $\widetilde{D}(R)$ decays as a power law (i.e.~$\widetilde{D}(R)\sim R^{-\eta}$ with $\eta>0$) due to order 
parameter phase or `spin wave'-like fluctuations \cite{SBanerjee2011_2}. We find here the 
consequences of these for the spectral function, i.e.~
\begin{eqnarray} 
A(\mathbf{k},\omega)&\equiv&-\frac{2}{\pi}\mathrm{Im}\left[G(\mathbf{k},i\omega_n\rightarrow\omega+i\delta)\right],
\label{eq.SpectralFunction}
\end{eqnarray}
 measured in ARPES. $G(\mathbf{k},i\omega_n)$ is obtained from
eq.\eqref{eq.DysonEquation} with the self energy calculated from eq.\eqref{eq.StaticSelfEnergy}
using the aforementioned forms of the pair propagator $\widetilde{D}(\mathbf{R})$ (or ${D}(\mathbf{R})$) for
temperatures above and below $T_c$.

 In a regime where the fluctuations in the real pair magnitude $\Delta_m$ are short ranged,
\begin{eqnarray}
D(\mathbf{R})\simeq \langle \Delta(\mathbf{R})\Delta(\mathbf{0})\rangle
\langle e^{i[\tilde{\phi}(\mathbf{R})-\tilde{\phi}(\mathbf{0})]}
\rangle\equiv
\bar{\Delta}^2 \bar{D}(R) \label{eq.PairCorrelator}
\end{eqnarray} 
for large $R$ ($R>>a$) [as evident from eq.\eqref{eq.UrsellFunction}, $\widetilde{D}(R)\simeq
\bar{\Delta}^2(\bar{D}(R)-|<e^{i\tilde{\phi}(\mathbf{0})}>|^2)$]. This 
decoupling between magnitude and long distance phase correlations is accurate for $x\aplt x_\mathrm{opt}$, a
manifestation of which is the separation between $T^*$ and $T_c$. Our calculations based on eq.\eqref{eq.PairCorrelator} 
are therefore reliable in this doping range. For strictly two-dimensional system mostly considered here, we use the general form \cite{PMChaikinBook} for the phase
correlator,
\begin{eqnarray}
\bar{D}(R)=(\tilde{\Lambda}R)^{-\eta}\exp{(-R/\xi)} \label{eq.QLRO_PhaseCorrelator}
\end{eqnarray}
 (with $\tilde{\Lambda}\sim a^{-1}$) in eq.\eqref{eq.StaticSelfEnergy}. An analytical expression can be obtained for the self energy in this case \cite{SBanerjee2011_2}, namely
\begin{eqnarray}
\Sigma(\mathbf{k},i\omega_n)&\simeq&\frac{-i~\mathrm{sgn}(\omega_n)\Gamma(1-\eta)\bar{\Delta}_\mathbf{k}^2}{\left(\tilde{\Lambda}\mathrm{v}_\mathbf{k}\right)^\eta\left(v_\mathbf{k}/\xi-i~\mathrm{sgn}(\omega_n)(i\omega_n+\xi_\mathbf{k})\right)^{1-\eta}}\label{eq.MatsubaraSelfEnergy}
\end{eqnarray}
In this equation, $\Gamma$ is the well known gamma function and
$\mathbf{v}_\mathbf{k}=\frac{1}{a}\frac{\partial\xi_\mathbf{k}}{\partial\mathbf{k}}$ (with
$v_\mathbf{k}=|\mathbf{v}_\mathbf{k}|$) is the velocity (expressed
in units of energy) obtainable from the energy dispersion $\xi_\mathbf{k}$.  The above self energy does not affect the 
nodal quasiparticles owing to the $\mathbf{k}$ dependence of 
$\bar{\Delta}_\mathbf{k}=(\bar{\Delta}/2)(\cos{k_xa}-\cos{k_ya})$ . We have also obtained electronic self-energy for an anisotropic three-dimensional system with small interlayer coupling, as appropriate for the cuprates \cite{SBanerjee2011_2}. We have used inputs from our phenomenological theory as well as from experimental data to estimate various parameters, such as $\bar{\Delta},~\xi,~\eta$, that enter into the self-energy \cite{SBanerjee2011_2}. We summarize the main results of our calculations for low-energy electronic spectral properties in the next section. 

\section{Results from the phenomenological theory and comparison with experiments} \label{Sec.Results}

\begin{figure} [hbt]
\begin{center}
\includegraphics[width=8cm,angle=-90]{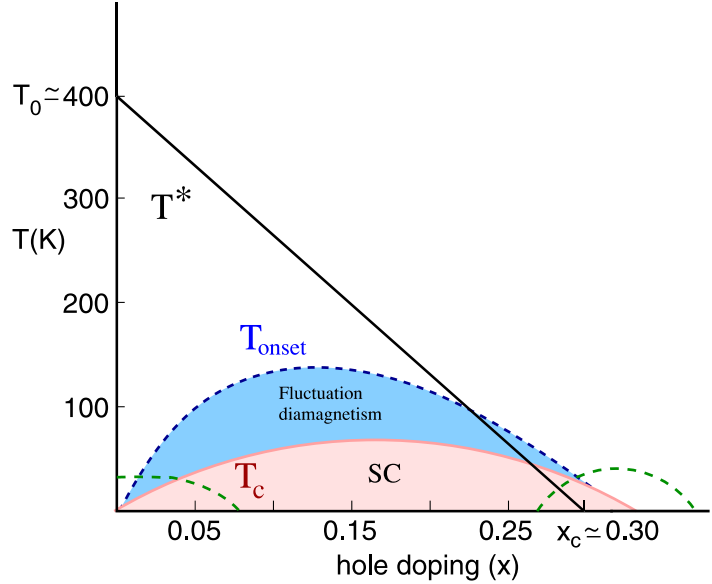}
\end{center}
\caption{\footnotesize{A schematic phase diagram for the model [eq.~\eqref{Eq.functional}] in the hole doping $x$ and temperature $T$ plane. The local pairing temperature scale $T^*(x)$,  (solid black line) is an input to our phenomenological model [eq.~\eqref{Eq.functional}] and it mimics the experimental pseudogap temperature scale. The model reproduces a dome-shaped superconducting region (drawn in pink). The region of enhanced fluctuation diamagnetism and Nernst signal (shaded in blue) in the pseudogap phase and corresponding onset temperature $T_\mathrm{onset}$ are shown. Between $T_c$ and $T^*$, electronic spectral functions shows an antinodal pseudogap. The pseudogap state is also characterized by Fermi arcs that emerges from $d$-wave node at $T_c$ and persist all the way up to $T^*$. The two arcs shown by dotted lines denote regions where quantum fluctuation effects, as well as other low-energy degrees of 
freedom, such as electronic and spin plus their coupling with pair degrees of freedom, need to be explicitly included in 
the free energy functional.}}
\label{fig.GLPhaseDiagram}
\end{figure}

 Fig.\ref{fig.GLPhaseDiagram} shows overall phase diagram obtained from our phenomenological theory and its extensions to capture thermoelectric transport and low-energy electronic spectra. We summarize our main results below and indicate how they compare with the cuprate phenomenology discussed in Section \ref{sec.cupratephenomena}.
 
 \begin{figure}[hbt]
\begin{center}
\begin{minipage}[c]{0.50\linewidth}
\includegraphics[width=\linewidth]{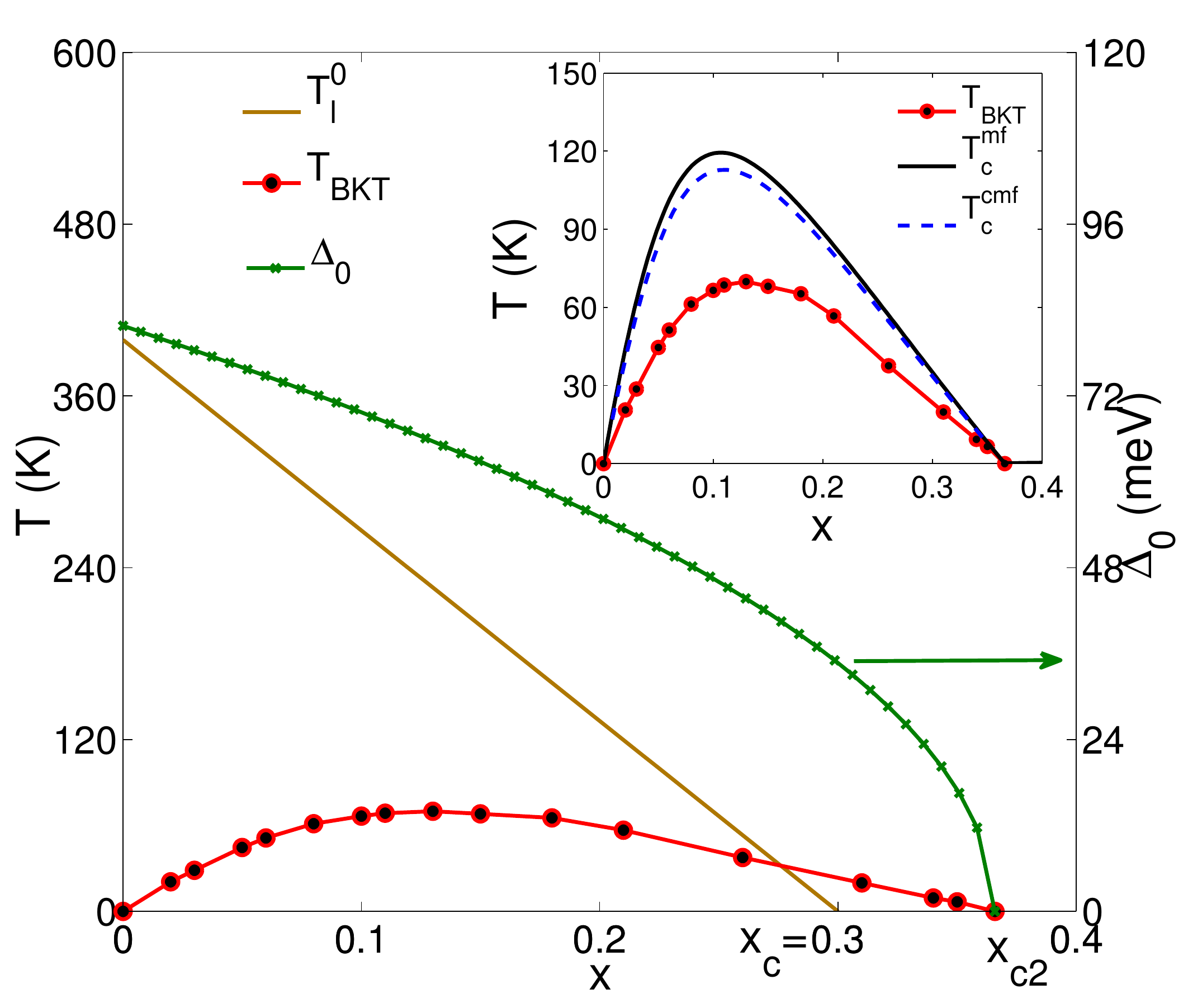}
\end{minipage}\hfill
\begin{minipage}[c]{0.38\linewidth}
\caption{{\footnotesize Doping dependence of different temperature scales ($T^*=T^0_l$ and
$T_\mathrm{BKT}$, the Berezinskii-Kosterlitz-Thouless transition temperature, as appropriate for the purely 2d system considered here) and the zero temperature gap $\Delta_0$ are shown in the main plot.}\label{fig.Tc}}
\end{minipage}
\end{center}
\end{figure}
 
\begin{figure}[hbt!]
\begin{center}
\includegraphics[height=6cm]{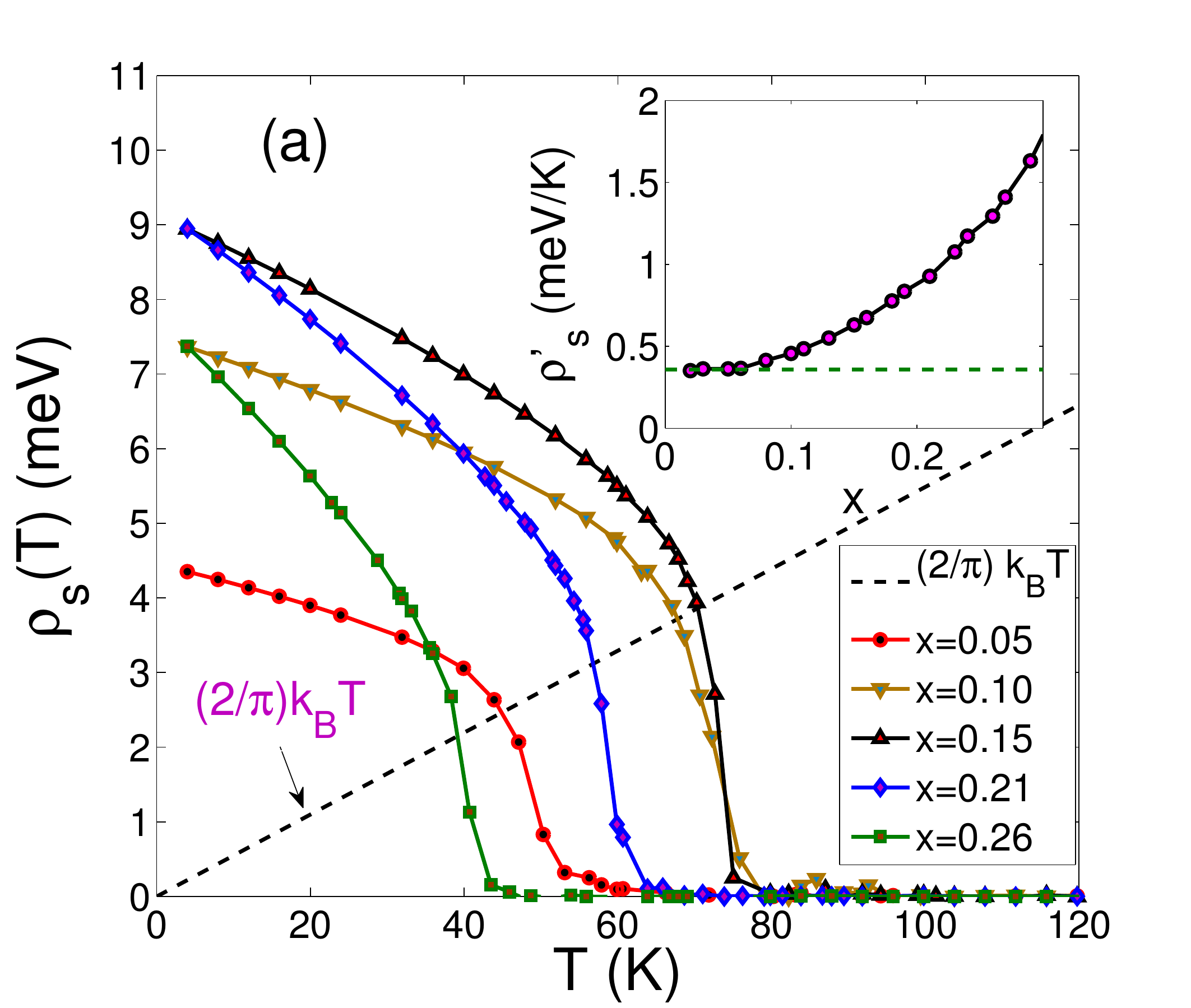}
\end{center}
\caption{\footnotesize Finite temperature superfluid density calculated via MC simulation for different $x$ values. 
The dashed line corresponds to the size of universal Nelson-Kosterlitz jump expected at a BKT
transition. $T_\mathrm{BKT}(x)$ has been obtained from the intersection of this line with $\rho_s(x,T)$ vs.~$T$ curves.} 
\label{fig.SuperfluidDensity}
\end{figure}

{\bf 1. Superconducting and pseudogap phases:} The pseudogap and superconducting states are both described by a non-zero thermal average of local pairing amplitude, i.e. $\Delta(T)=\langle \Delta_m\rangle\neq 0$. The superconducting state has long-range phase coherence characterized by superfluid density $\rho_s(x,T)$ becoming non-zero below a transition temperature $T_c(x)$ (Fig.\ref{fig.GLPhaseDiagram}), that reproduces the nearly parabolic superconducting dome of Fig.\ref{fig.CupratePhaseDiagram}. The results for the zero-temperature gap, $T_c$ as a function of $x$ and superfluid density as function of $x$ and $T$ are shown in Figs.\ref{fig.Tc},\ref{fig.SuperfluidDensity}. We have also calculated the effect of quantum phase fluctuations on $T_c$ in the underdoped regime; $T_c$ is pushed to zero at nonzero $x$, leading to a $T_c(x)$ dome \cite{SBanerjee2011_1} in close agreement with experiment.

\begin{figure}
\begin{center}
\includegraphics[height=6cm]{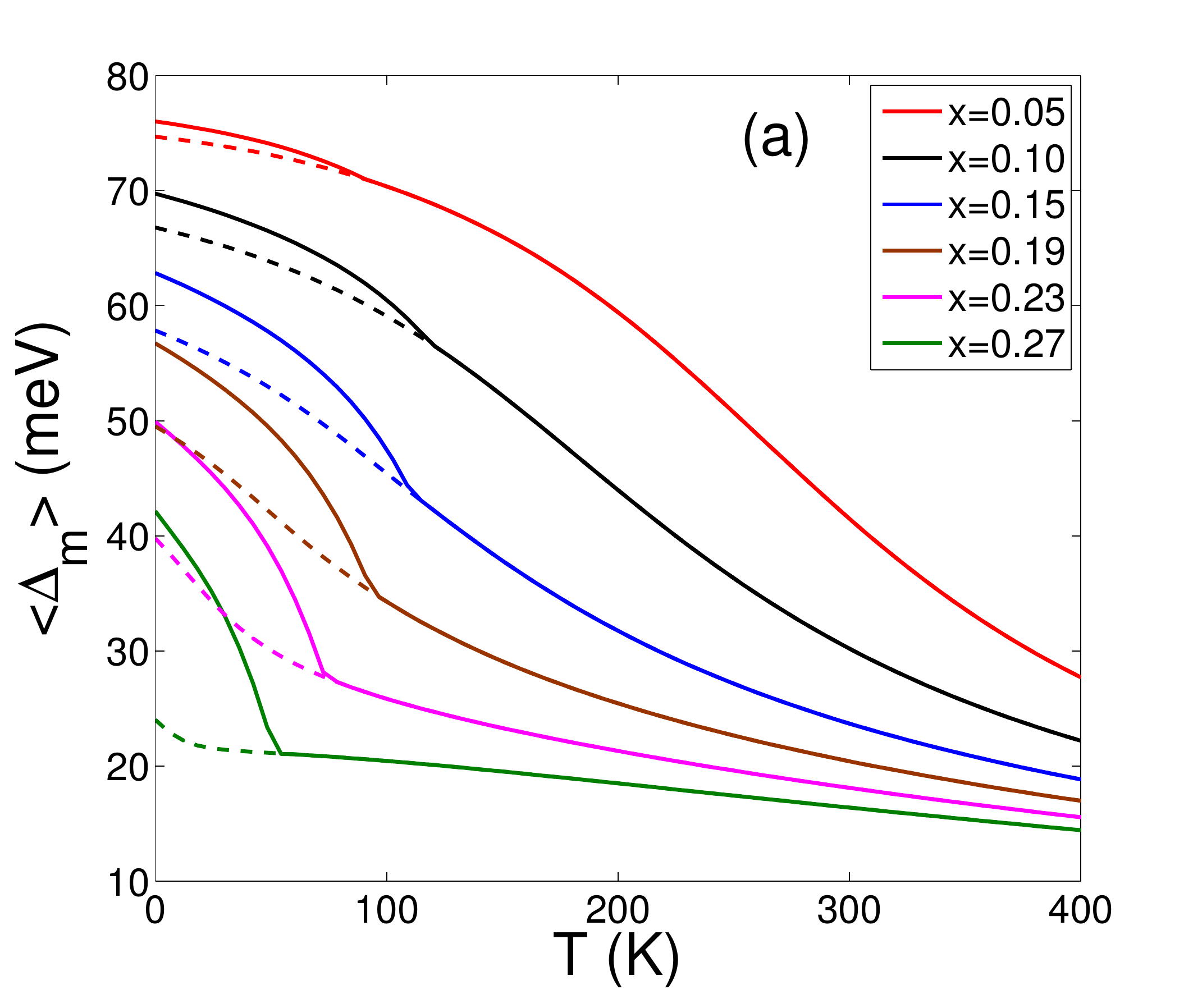}
\end{center}
\caption{{\footnotesize The figure shows a crossover $\langle \Delta_m\rangle$ from small to large values around $T^*(x)$ and an onset of the second gap feature at $T_c$.}}
\label{fig.Gap}
\end{figure}

{\bf 2. Pseudogap crossover:} The onset of pesudogap state is characterized by rapid but smooth crossover of various quantities across $T^*(x)$ which is an input incorporated via change of sign of the coefficient $A$ [eq.\eqref{Eq.functional}] in our theory. For example, the local pairing amplitude $\Delta(T)$ rises rapidly with decreasing temperature across $T^*(x)$ and the pairing contribution to the specific heat has a broad `hump' around the same temperature \cite{SBanerjee2011_1}. Both of these are consistent with experiments, namely spin gap seen in Knight shift (Section \ref{Sec.PseudogapPhase}) and broad feature seen in specific heat measurements \cite{Matsuzaki2004}. In our theory, a similar crossover also appears at a temperature where the antinodal pseudogap in electronic spectral function is completely filled up or, in other words, the Fermi arc extends over the full Fermi surface \cite{SBanerjee2011_2}. All these various crossovers appears around temperature $T^*(x)$, however not all at exactly same temperature but over a somewhat broad range around $T^*(x)$ \cite{SBanerjee2011_1}(see Fig.\ref{fig.GLPhaseDiagram}), justifying the pseudogap temperature scale as a crossover rather than a sharp transition, as seen in the experiments \ref{Sec.PseudogapPhase}. The results for pseudogap crossover seen in local pairing amplitude and specific heat are shown in Figs.\ref{fig.Gap},\ref{fig.SpecificHeat}.

\begin{figure}[hbt!]
\begin{center}
\includegraphics[height=6cm]{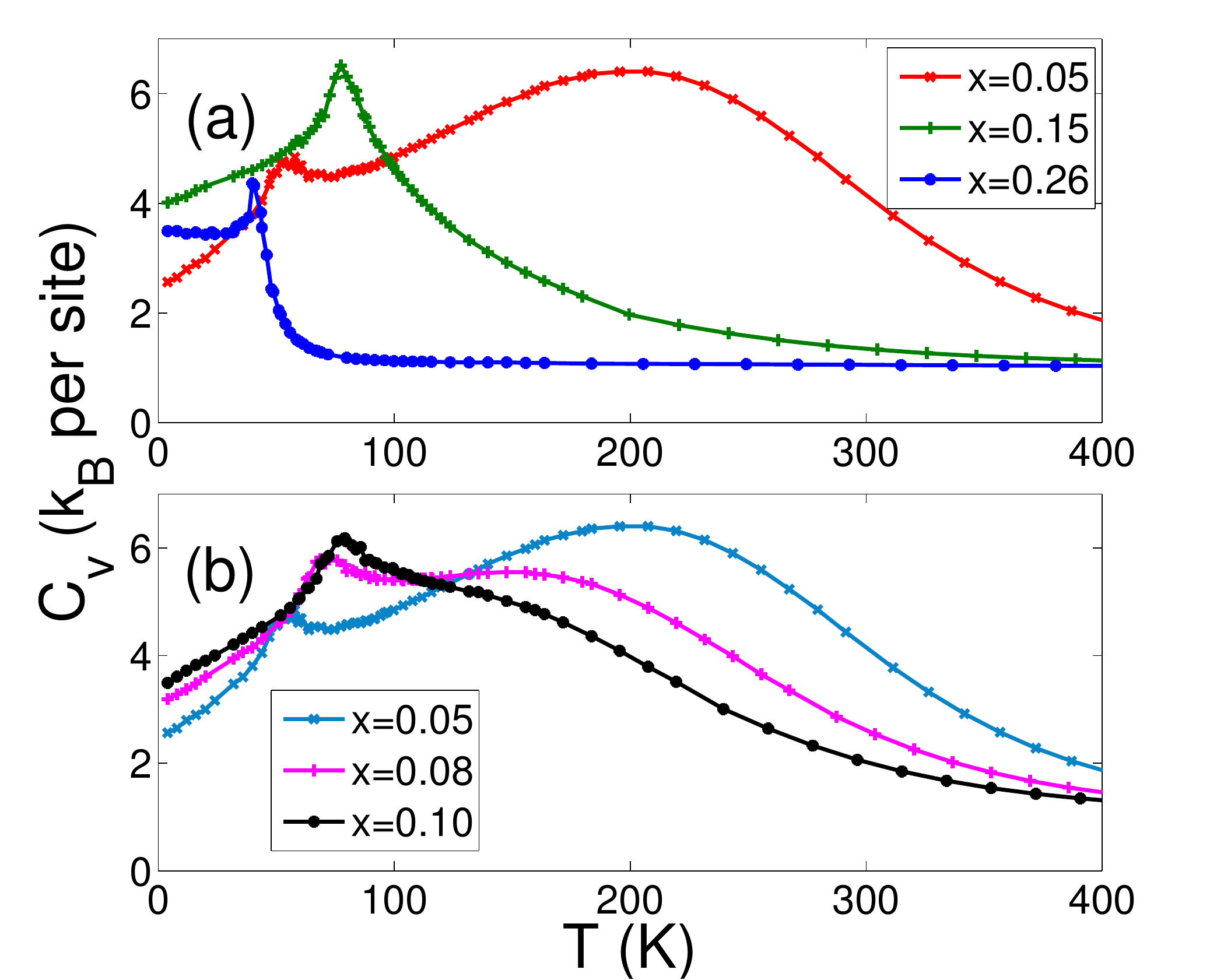}
\end{center}
\caption{\footnotesize (a) Specific heat obtained from MC simulation of our model. Panel (b) shows the evolution of the broad maximum around $T^*$ with doping in the
underdoped region.}
\label{fig.SpecificHeat}
\end{figure}

\begin{figure}[H]
\centering
\includegraphics[height=6cm]{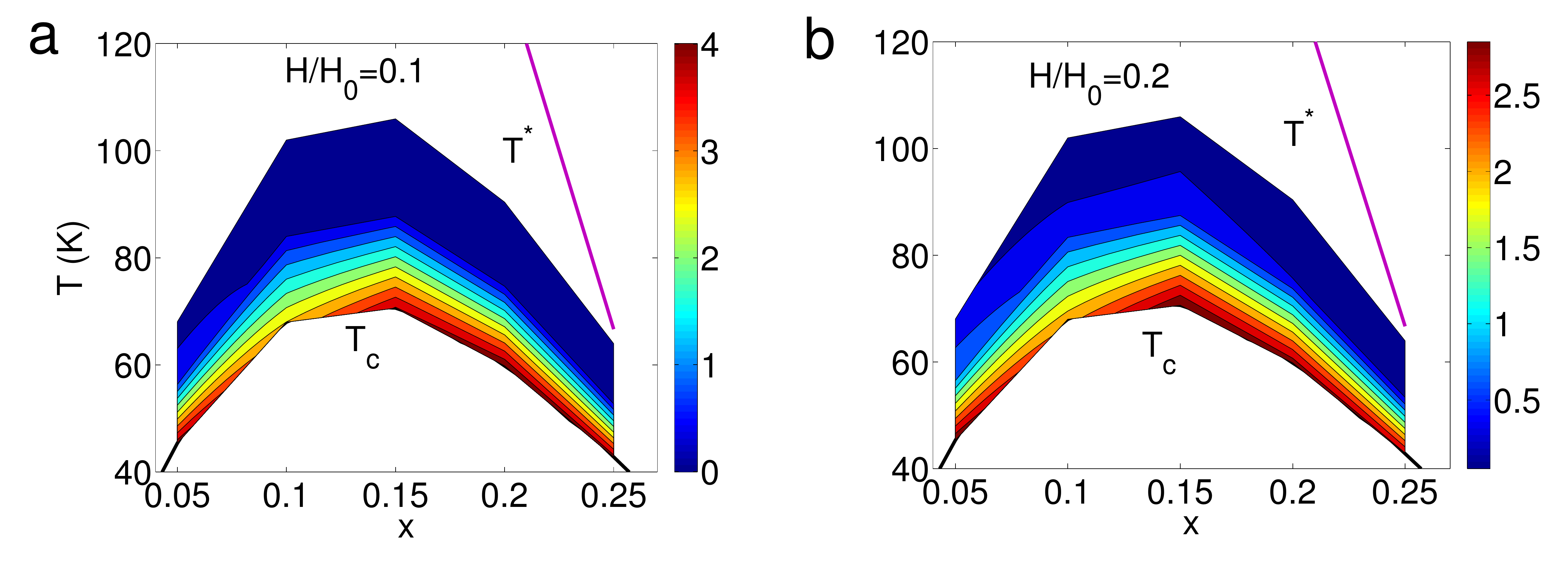}
\caption{$\alpha_{xy}$ in the $x-T$ plane for two different values of the magnetic field (a) $H/H_0=0.1$ and (b) $H/H_0=0.2$. The lines of constant $\alpha_{xy}$ follow the superconducting dome. This indicates that the equilibrium superconducting fluctuations responsible for the suppression of the superfluid stiffness also determine the thermoelectric response.}\label{fig:Nern_PD}
\end{figure}

{\bf 3. Large region of fluctuation diamagnetism and Nernst effect above $T_c$:} We obtain a region of enhanced fluctuation diamagnetism in the pseudogap phase extending far above (around 1.5 times) the transition temperature $T_c$ and show that the boundary of the region, namely the onset temperature $T_{onset}(x)$, follows a dome shaped curve tracking $T_c$ as a function of doping (Fig.\ref{fig.GLPhaseDiagram}). The results for Nernst regime computed from our theory is shown in Fig.\ref{fig:Nern_PD}. Similar feature is also seen for magnetization \cite{KSarkar2016}. 

Experimentally, both the Nernst effect~\cite{Wang2006} and diamagnetism~\cite{Li2010} have been seen to track the superconducting dome. As mentioned earlier, the persistence of the Nernst and diamagnetism signal over a dome-shaped region above $T_c$, instead of the entire pseudogap state till $T^*(x)$, has been argued as evidence against the pairing origin of the pseudogap line. This is due to the expectation that if the pseudogap line is related to pairing then superconducting fluctuations should continue till $T^*$. However, on the basis of our results, we can argue that this expectation is not justified since pairing fluctuations identifiable as superconducting fluctuations, e.g.~those detected through Nernst or diamagnetism, are mainly controlled by $\rho_s$ or the $T_c$ scale. The GL-like model we study has the pseudogap temperature $T^*(x)$ explicitly set as the local pairing scale by construction, but, even then, the diamagnetic signal tracks the superconducting $T_c$ that is governed by the superfluid density rather than the pairing scale $T^*$ on the underdoped side. The suggestion is that the long wavelength anomalies in Nernst effect and fluctuation diamagnetism require a large correlation length for superconducting phase fluctuations as happens near $T_c$.

\begin{figure} [hbt]
\begin{center}
\begin{minipage}[c]{0.68\linewidth}
\includegraphics[width=\linewidth]{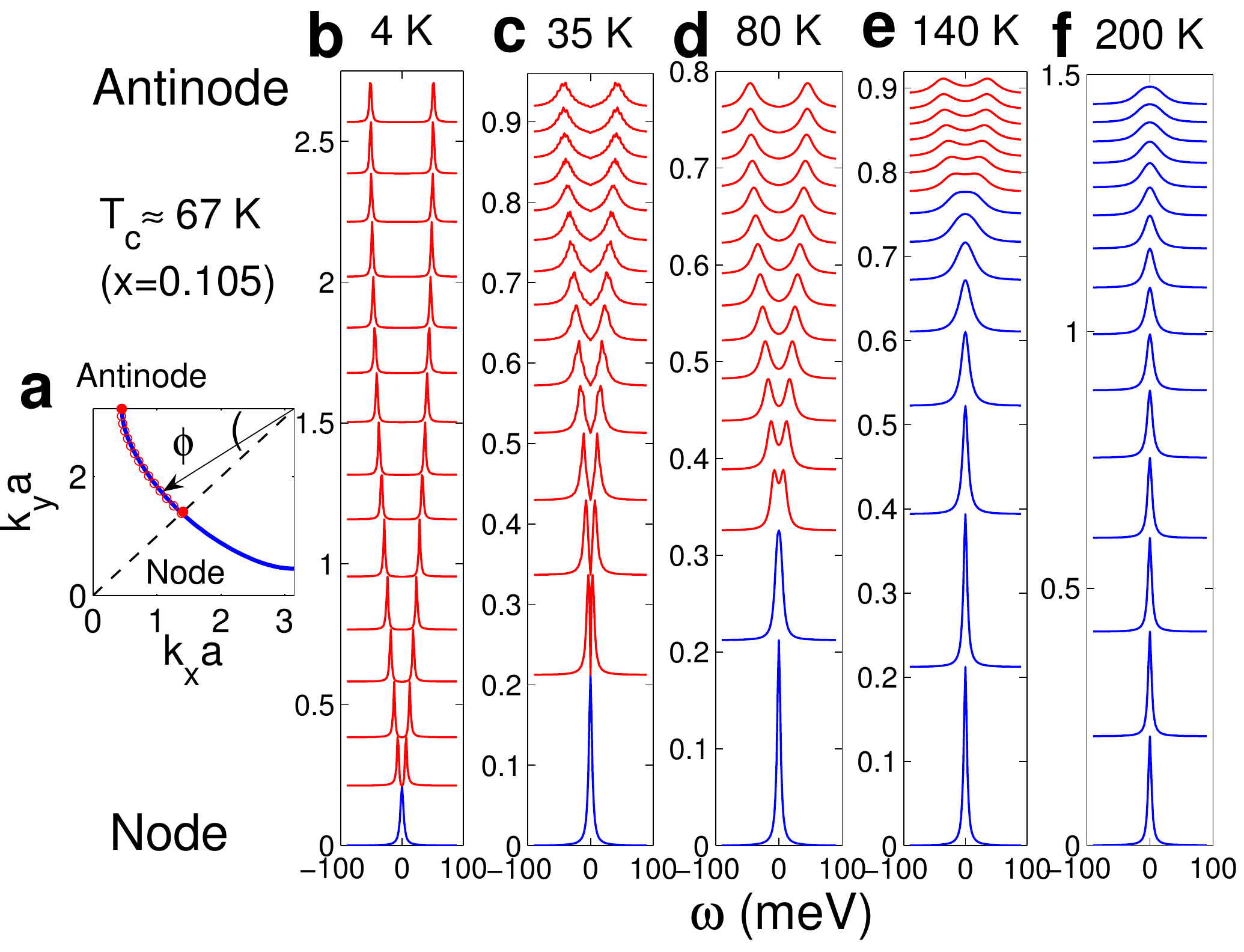}
\end{minipage}\hfill
\begin{minipage}[c]{0.28\linewidth}
\caption{{\footnotesize {\bf The Fermi surface and the spectral density along it.} {\bf a}, Fermi surface (FS) in the first
quadrant of the Brillouin zone for $x\simeq 0.11$, a typical value. The angular position $\phi$ on the Fermi
surface is shown. {\bf b-f}, Single-particle spectral density $A(\mathbf{k},\omega)$ at 15 equally spaced points 
(red circles) on the Fermi surface in {\bf a} shown vertically shifted from node (bottom) to antinode (top), as a 
function of temperature at $x\simeq 0.11$. Blue and red lines correspond to Fermi arc and gapped portions of FS,
respectively. (From Ref.\onlinecite{SBanerjee2011_1})}\label{fig.EDC}}
\end{minipage}
\end{center}
\end{figure}

\begin{figure*} [hbt]
\begin{center}
\includegraphics[height=7cm]{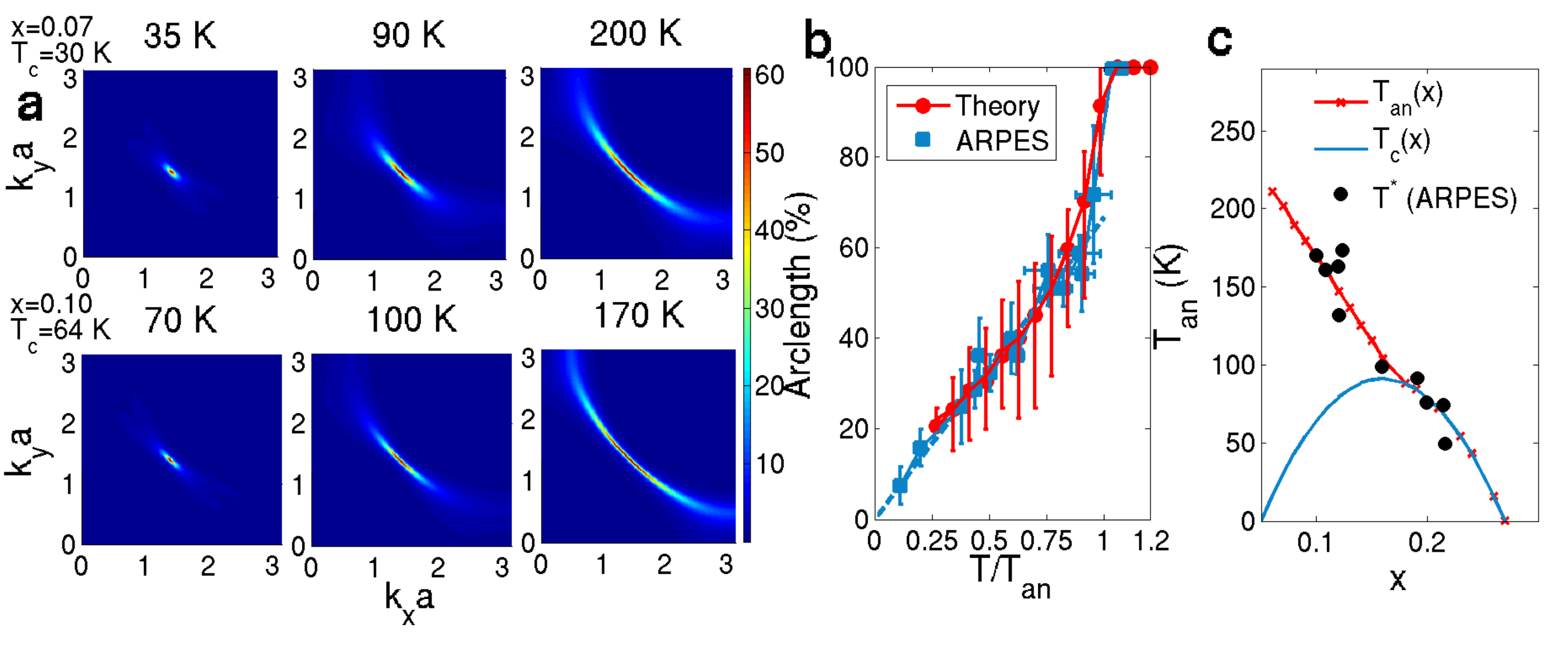}
\end{center}
\caption{{\footnotesize{\bf Fermi arcs above $T_c$.} {\bf a}, Colourmap of
$A(\mathbf{k},\omega=0)$ over the first quadrant of the Brillouin zone for two values of $x$ at
different temperatures. The Fermi arc is easily picked out visually. The color bar indicates the value of 
$A(\mathbf{k},\omega=0)$. {\bf b}, The 
arc length vs. $T/T_\mathrm{an}$ curve, averaged over the 
entire doping range ($x\aplt x_\mathrm{opt}$). On the $y$-axis, $0\%$ is the node and $100\%$ is the antinode. The 
experimental data are from Ref.\onlinecite{AKanigel2006}. The vertical error bars in the
theoretical points indicate the variation of the arc length at different $x$ for a given $T/T_\mathrm{an}(x)$.
{\bf c}, The antinodal pseudogap filling temperature $T_\mathrm{an}$ as a function of $x$ is compared
with the data for the pseudogap temperature $T^*$ from ARPES \cite{AKanigel1}. (From Ref.\onlinecite{SBanerjee2011_2})}}
\label{fig.FermiArc}
\end{figure*}

{\bf 4. Filling up of pseudogap and evolution of Fermi arc with doping and temperature:} We show that, in remarkable agreement with ARPES experiments, non-zero local pairing amplitude above $T_c$ manifests itself as pseudogap in the electronic spectral function on the \emph{putative} Fermi surface that exists above the pseudogap temperature scale. Two spectral peaks appear away from the $\omega=0$, however their is finite spectral weight at the zero energy. Above $T_c$, finite near-nodal gapless regions on the Fermi surface, or Fermi arcs, emerges from the $d$-wave node below $T_c$. The arcs expands with increasing temperatures and finally take over the whole Fermi surface at a temperature $T_\mathrm{an}\approx T^*(x)$, where the antinodal pseudogap gets completely filled up. The results are shown in Figs.\ref{fig.EDC},\ref{fig.FermiArc}.

{\bf 5. Deviation from $d$-wave gap structure below $T_c$ due to phase fluctuations: `Two gaps' from one gap:} 
\begin{figure}[hbt]
\begin{center}
\includegraphics[height=8cm]{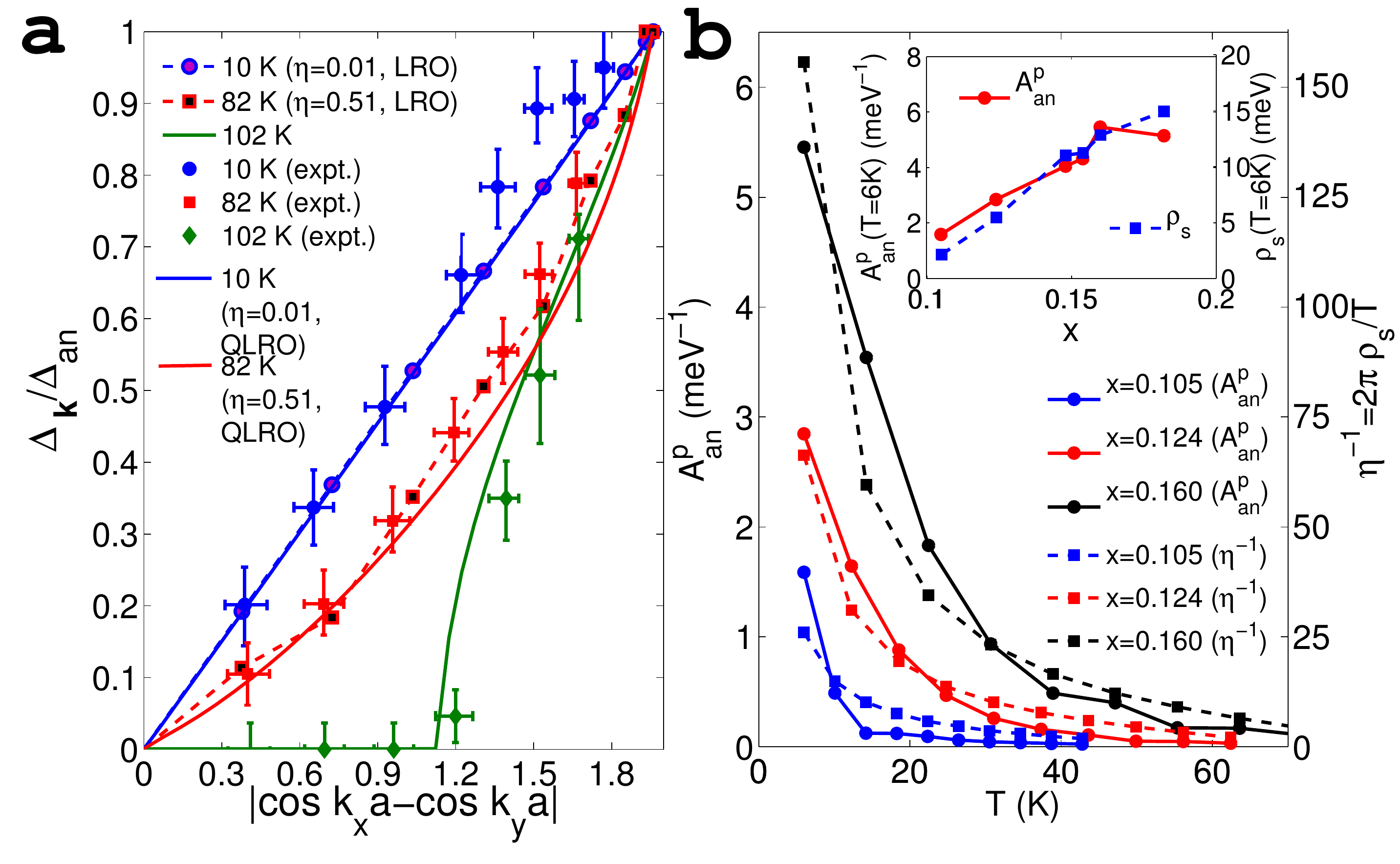}
\end{center}
\caption{{\footnotesize{\bf Spectral properties below $T_c$.} {\bf a}, Deviation of $\Delta_\mathbf{k}$ from the $d$-wave 
form, $|\cos{k_xa}-\cos{k_ya}|$, below $T_c$ for $x=0.16$ is compared with $\Delta_\mathbf{k}$ obtained in 
ARPES \cite{WSLee2007}. At low temperature ($T=10$ K), $\Delta_\mathbf{k}$ follows the canonical $d$-wave form 
(a straight line), but bends away 
considerably from it close to $T_c$. For $T<T_c$, we have shown $\Delta_\mathbf{k}$
obtained for both the true LRO phase [i.e.~$D(R\rightarrow \infty)=|\langle\psi_m\rangle|^2\neq 0$] and the quasi-LRO or 
QLRO phase [i.e.~$D(R)\sim R^{-\eta}\xrightarrow{R\rightarrow \infty}0$]. {\bf b}, The antinodal peak height
$A^p_\mathrm{an}$ vs.~$T$ tracks $\rho_s(T)/T$ vs.~$T$. {\bf Inset} At low temperatures ($T=6$ K) $A^p_\mathrm{an}(x)$ 
closely follows $\rho_s(x)$, especially in the underdoped side ($x\aplt 0.16$).}}
\label{fig.PeakStrength}
\end{figure}

                   Possibly the most widely explored property of a superconductor is the energy gap
$\Delta_\mathbf{k}$ or $\Delta(\phi)$. The apparent deviation (`bending') of the gap below $T_c$, 
inferred from ARPES \cite{WSLee2007}, from the $d$-wave form has been the subject of a great deal of current 
interest, leading to speculation that there are two gaps in high-$T_c$ superconductors \cite{AJMillis2006}. We have obtained 
the gap  $\Delta_\mathbf{k}$ from the peak position of the calculated spectral function $A(\mathbf{k},\omega)$ for a fixed 
$\mathbf{k}$ as mentioned earlier and conclude from the results 
(see Fig.\ref{fig.PeakStrength}{\bf a}) that the `bending' is due to the coupling of the electron to thermal phase 
fluctuations (`spin waves') below $T_c$. As expected from such an origin, it is large close to $T_c$, and small as
$T\rightarrow 0$. These results confirm that there is only one gap, but that to `uncover' it, effects of
coupling to pair fluctuations (`spin waves') have to be factored in.

\section{Conclusion and future directions}\label{sec.conclusion}
The cuprate superconductors are unusual materials both in their superconducting and non-superconducting states. After a description of some of their strangeness (Section \ref{sec.cupratephenomena}), we have described a phenomenological theory that focuses on their superconductivity and on the properties connected with pairing fluctuations. We have also described a simple model which couples the fluctuations to electrons. Our aim has been to show that such an approach, in principle and in practice, successfully makes sense of the large amount of experimental information accumulated over decades on these materials, and that it does this both qualitatively and quantitatively. Given that such an approach is made in the background of our already strong and increasing knowledge of electrons in condensed matter, and given the repeated discovery of major novel experimental facts, some directions suggest themselves, in addition to several extensions, such as the inclusion of quantum phase fluctuations in the GL theory, already mentioned. One direction, within the ambit of phenomenological theory, is to enlarge it to include competing order. In the last several years, CDW (charge density wave) correlations have been clearly seen in these materials \cite{Fradkin2015,Ghiringhelli2012,Chang2012,Achkar2012}. Their dependence on doping, temperature and magnetic field \cite{Ghiringhelli2012,Chang2012,Achkar2012,TWu2011,DLeBoeuf2013,SGerber2015}, whether the order is short or long range \cite{Ghiringhelli2012,Chang2012,Achkar2012}, whether they are bond CDW's or not \cite{Hamidian2016}, have all been documented extensively. Whether this is a competing order and whether there is ineluctable complexity in these compounds, are important questions. There is again considerable experimental support for other non Cooper pair correlations such as electronic liquid crystal like effects \cite{Kivelson1998,Fradkin2015}, SDW (spin density wave) like correlations \cite{Kivelson2003}. 

The approach emphasizes the importance of phase fluctuations and is really a theory in which Cooper pair magnitude $\Delta_m$ and the phase $\phi_m$ are both relevant low-energy degrees of freedom whose significance varies with doping $x$ and temperature $T$. For small $x$, the amplitude fluctuations are not important. This is like in interacting spin models where spin length fluctuations are not important, only there directional fluctuations which are like phase angle ($\phi_m$) fluctuations are thermally excited. As $x$ increases, amplitude $\Delta_m$ fluctuations have decreasing energies, till presumably for large $x$, one has the BCS like limit where both these are indistinguishable energetically. This raises several important questions. One is that there are and must be other phase angle fluctuations besides the $d$-wave like ones in which we constrain the nearest neighbor $x,y$-bond angles to be $(\pi/2)$ out of phase. These have been investigated recently \cite{} and have important consequences. The other is whether there is indeed a BCS limit, given the short coherence length. Questions connected with time-reversal symmetry breaking \cite{Varma2006,Fauque2006}, anomalous Kerr effect \cite{Xia2008} etc. are other basic directions. Can the effects be understood in the above framework? 

An overarching issue is the need for a microscopic theory with \emph{electrons} as basic ingredients. The analogue is the celebrated BCS theory which followed the original GL theory in seven or so years. Here, nearly a quarter of century after the discovery of cuprate superconductivity, and in the background of a large number of well developed many-body electronic approaches, a phenomenological theory has been proposed. The underlying question  is the well known problem of a theoretical description of strongly correlated mobile electrons. Many phenomena of the strange metal phase itself, depend for their theoretical understanding on the development of such a theory, so that the development of a `reliable' many-particle approach for strongly correlated electrons with inevitable pairing correlations is a pressing necessity for a comprehensive theory of electronic matter.inductable
\appendix

\section{Parameters of the phenomenological theory}\label{app.GLParameters}

As natural in a phenomenological theory, the parameters of the functional in eq.\eqref{Eq.functional} are chosen to be consistent with experiment. The doping and temperature dependence of the coefficients are parametrized as $A(x,T)= (f/T_0)^2[T-T^*(x)]e^{T/T_0}$, $B=bf^4/T_0^3$ and $C(x)=xcf^2/T_0$; $f$, $b$, $c$ are dimensionless and $T^*(x)=T_0(1-x/x_c)$ with the single energy scale $T_0$ and doping concentration $x_c=0.3$ controlling the pseudogap temperature scale. The phenomenological parameters $f$, $b$, $c$ vary for different cuprates and $T_0$ is the bare pseudogap temperature extrapolated to zero doping. The exponential factor $e^{T/T_0}$ appearing in $A$ is not very crucial in the temperature range of our interest, i.e. below the pseudogap temperature. This factor suppresses average local gap magnitude $\langle \Delta_m\rangle$ at high temperatures ($T\apgt T^*(x)$) with respect to its otherwise temperature independent equipartition value $\sqrt{T/A(x,T)}$ which will result from the simplified form of the functional [Eq.\eqref{Eq.functional}] being used over the entire range of temperature. Such a suppression is natural in a degenerate Fermi system; the relevant local electron pair susceptibility is rather small above the pair binding temperature and below the degeneracy temperature. 

The forms of the parameters $A$, $B$ and $C$ specified above allow us to reproduce the superconducting dome. However, it is also important to mention that having chosen $A$, $B$ and $C$ to reproduce the superconducting dome, the doping and temperature dependences of other physical properties like
the superfluid density, the magnitude of the local gap, specific heat, orbital magnetization and Nernst coefficient also agree
very well with experiments~\cite{SBanerjee2011_1,KSarkar2016,KSarkar2017}. 

While, underdoped phenomenology plays an important part in determining the form of our model, e.g.~to determine the doping dependence of the parameter $C$ as mentioned above, it also produces the standard GL theory for conventional superconductors on the overdoped side. The amplitude of pairing approaches zero at $T_c$, or in other words the actual or `renormalized' pairing scale $\widetilde{T}^{*}(x)\approx T_c(x)$, on the overdoped side in our theory \cite{SBanerjee2011_1}. This is in conformity with the common expectation that the BCS theory or mean-field GL theory is more appropriate for overdoped cuprates.

For specific values of the parameters, e.g.~for Bi2212, which has a $T_c^\mathrm{opt}\simeq 91$ K at $x=x_\mathrm{opt}\simeq 0.15$, we choose $f\simeq 1.33$, $b=0.1$, $c\simeq0.3$ with $T_0\simeq 400$ K. This choice of parameters leads to an optimal BKT transition temperature $T_\mathrm{KT}^{\mathrm{opt}}\approx 75 K$ for the 2D system that we study. In the cuprates, the small but finite inter-layer coupling between $\mathrm{CuO}_2$ planes is expected to lead to a slightly higher $T_c$. The interlayer coupling can be easily incorporated in our model in the manner of Lawrence and Doniach\cite{Lawrence1970}. Since this coupling is, in practice, quite small (e.g.~the measured anisotropy ratio in Bi2212 is about 100), it makes very little difference quantitatively to most of our estimates.
\bibliography{HighTc}

\end{document}